\documentclass{acm_proc_article-sp}
\usepackage{amsmath}
\usepackage{stmaryrd}
\usepackage{mathbbold}
\usepackage{amsfonts}
\usepackage{txfonts}
\usepackage{amssymb}
\usepackage{balance}
\usepackage{times}
\usepackage{helvet}
\usepackage{courier}
\usepackage{graphicx}
\usepackage{subfigure}
\usepackage{multirow}
\usepackage{algorithmic}
\usepackage{algorithm}
\usepackage{mathcomp}

\newtheorem{problem}{Problem}
\newtheorem{lemma}{Lemma}
\newcommand{\mat}[1]{{\bf #1}}   

\newcommand{\hide}[1]{}
\newcommand{\QED}{\hfill $\Box$ \hfill}
\newcommand{\cop}{{\em CoPs}}
\newcommand{\copit}{{\em CoPs-Iter}}
\newcommand{\copqg}{{\em CoPs-QG}}
\newcommand{\copqq}{{\em CoPs-QQ}}
\newcommand{\copgg}{{\em CoPs-GG}}
\newcommand{\copgq}{{\em CoPs-GQ}}

\begin{document}

\title{Want a Good Answer? Ask a Good Question First!}

\numberofauthors{5}

\author{
\alignauthor Yuan Yao\\
       \affaddr{State Key Laboratory for Novel Software Technology, China}\\
       \email{yyao@smail.nju.edu.cn}
\alignauthor Hanghang Tong\\
       \affaddr{City College, CUNY, USA}\\
       \email{tong@cs.ccny.cuny.edu}
\alignauthor Tao Xie\\
       \affaddr{University of Illinois at Urbana-Champaign, USA}\\
       \email{taoxie@illinois.edu}
\and
\alignauthor Leman Akoglu\\
       \affaddr{Stony Brook University, USA}\\
       \email{leman@cs.stonybrook.edu}
\alignauthor Feng Xu\\
       \affaddr{State Key Laboratory for Novel Software Technology, China}\\
       \email{xf@nju.edu.cn}
\alignauthor Jian Lu\\
       \affaddr{State Key Laboratory for Novel Software Technology, China}\\
       \email{lj@nju.edu.cn}
}

\maketitle

\begin{abstract}
Community Question Answering (CQA) websites have become valuable repositories which host a massive volume of human knowledge. To maximize the utility of such knowledge, it is essential to evaluate the quality of an existing question or answer, especially soon after it is posted on the CQA website.

In this paper, we study the problem of inferring the quality of questions and answers through a case study of a software CQA (Stack Overflow). Our key finding is that the quality of an answer is strongly positively correlated with that of its question. Armed with this observation, we propose a family of algorithms to {\em jointly} predict the quality of questions and answers, for both quantifying numerical quality scores and differentiating the high-quality questions/answers from those of low quality. We conduct extensive experimental evaluations to demonstrate the effectiveness and efficiency of our methods.
\end{abstract}




\section{Introduction}
Community Question Answering (CQA) websites have become valuable repositories which host a massive volume of human knowledge. In addition to providing answers to the questioner, CQA websites now serve as knowledge bases for the searching and browsing of a much larger audience. For example, in the software forum Stack Overflow, programmers could post their programming questions on the forum, and others could propose their answers for these questions. Such questions as well as their associated answers could be valuable and reusable for many other programmers who encounter similar problems. In fact, millions of programmers now use Stack Overflow to search for high-quality answers to their programming questions, and Stack Overflow has also become a knowledge base for people to learn programming skills by browsing high-quality questions and answers~\cite{osbourn2011getting}.

To maximize the utility of CQA websites, predicting the quality of existing questions and answers, especially soon after they are posted, becomes an essential task for both information producers and consumers. From the perspective of information producer (e.g., who asks or answers questions), predicting the quality as early as possible could help the questions that are potentially of high quality to attract more high-quality answers by recommending these questions to experts. From the perspective of information consumer (e.g., who searches or browses questions and answers), it would be helpful to highlight the question-answer pairs with high quality (e.g., by displaying them more prominently on the website or allowing the search engine to be aware of their quality), so that users can easily discover them.
In addition to recognizing high-quality posts, quality prediction is also useful to detect useless posts or even spams. For example, Stack Overflow has launched a contest to predict whether a question will be closed due to quality problems at the moment the question is posted~\footnote{http://blog.stackoverflow.com/2012/08/stack-exchange-machine-learning-contest/}.

To date, a lot of efforts have been made to study the quality prediction problem in CQA websites. However, most of them treat questions and answers {\em separately} (See Section 5 for a review). For example, some work evaluates the question quality to recommend questions or to better match the user's query (e.g.~\cite{song2008question,sun2009learning}), while some work evaluates the answer quality to re-rank the returned answers from the search engine (e.g.~\cite{jeon2006framework,suryanto2009quality}).

We conjecture that there might be {\em correlation} between the quality of a question and that of its associated answers. Intuitively, an interesting question might obtain more attention from potential answerers and thus has a better chance to receive high-quality answers. On the other hand, it might be very difficult for a low-quality question to attract a high-quality answer due to, e.g., its poor expression in language, lack of interest in topics, etc.

Starting from this conjecture, we study in this paper the relationship between the quality of questions and that of answers through a case study of a software CQA (Stack Overflow). Our key finding is that the quality of an answer is indeed strongly positively correlated with that of its question. Armed with this observation, we propose a family of algorithms (\cop) to {\em jointly} predict the quality of questions and answers. The proposed \cop\ algorithms can be applied to both the continuous case (e.g., to infer numerical quality scores) and the binary case (e.g., to differentiate the high quality questions/answers from those of low quality). We conduct extensive experimental evaluations to demonstrate the effectiveness and efficiency of our methods. By collectively exploring the features, the predicted quality, and the link between them for both questions and answers simultaneously, our method achieves up to 13.13\% improvement over the state-of-the-art methods wrt prediction error. In addition, our method scales linearly wrt the number of questions and answers.



In summary, this paper makes the following three main contributions:
\begin{itemize}
  \item {\em Findings}. We empirically study the post quality in Stack Overflow. To the best of our knowledge, we are the first to quantitatively validate the correlation between the question quality and its associated answer quality.
  \item {\em Algorithms}. We propose a family of co-prediction algorithms \cop\ to jointly predict the quality of questions and answers for both quantifying numerical quality scores and differentiating the high-quality questions/answers from those of low quality .
  \item {\em Evaluations}. Extensive experimental evaluations on the Stack Overflow dataset demonstrate the effectiveness and efficiency of our methods.
\end{itemize}

The rest of the paper is organized as follows. Section 2 empirically studies the Stack Overflow dataset. Section 3 and 4 present the problem definitions and the proposed algorithms for the question/answer quality co-prediction problem, respectively. Section 5 presents the experimental results. Section 6 reviews related work. Section 7 discusses the future work, and Section 8 concludes the paper.

\section{Dataset Description and Analysis}
In this section, we describe the Stack Overflow dataset and our pre-processing steps on the dataset, followed by an empirically study of the question/answer quality in Stack Overflow.

\subsection{Dataset Description and Pre-processing}

Stack Overflow, which is ranked $3^{rd}$ among reference websites and $79^{th}$ among all websites~\footnote{http://www.alexa.com/}, is a popular CQA for software developers to ask and answer programming questions. The Stack Overflow dataset that we use spans from July 31, 2008 to August 31, 2011, and it is officially published and publicly available~\footnote{http://blog.stackoverflow.com/category/cc-wiki-dump/}.

We performed the following pre-processing steps on the raw data. First of all, we clean up the dataset by deleting the data where some important attributes are missing (e.g., some answers do not indicate which questions they belong to). After this step, the dataset contains 756,695 users, 1,966,272 questions, 4,282,570 answers, 14,056,000 votes, 1,055,500 favorites, and 7,306,298 comments. In the existing questions, 1,815,140 (92.3\%) of them have at least one answer. We only use these answered questions because our focus is on the quality correlation between questions and their answers. Next, as our primary goal is to predict the quality in the early stage of the questions/answers, we only consider the available information in the first 24 hours after the question is posted, and this step results in 3,577,041 answers (which means 83.5\% of the answers arrive in the first 24 hours). Our analysis is based on these 1,815,140 questions and their 3,577,041 associated answers.

\subsection{Empirical Findings}

\begin{figure}[t]
  \centering
  \subfigure[Questions]{
    \includegraphics[width=1.5in]{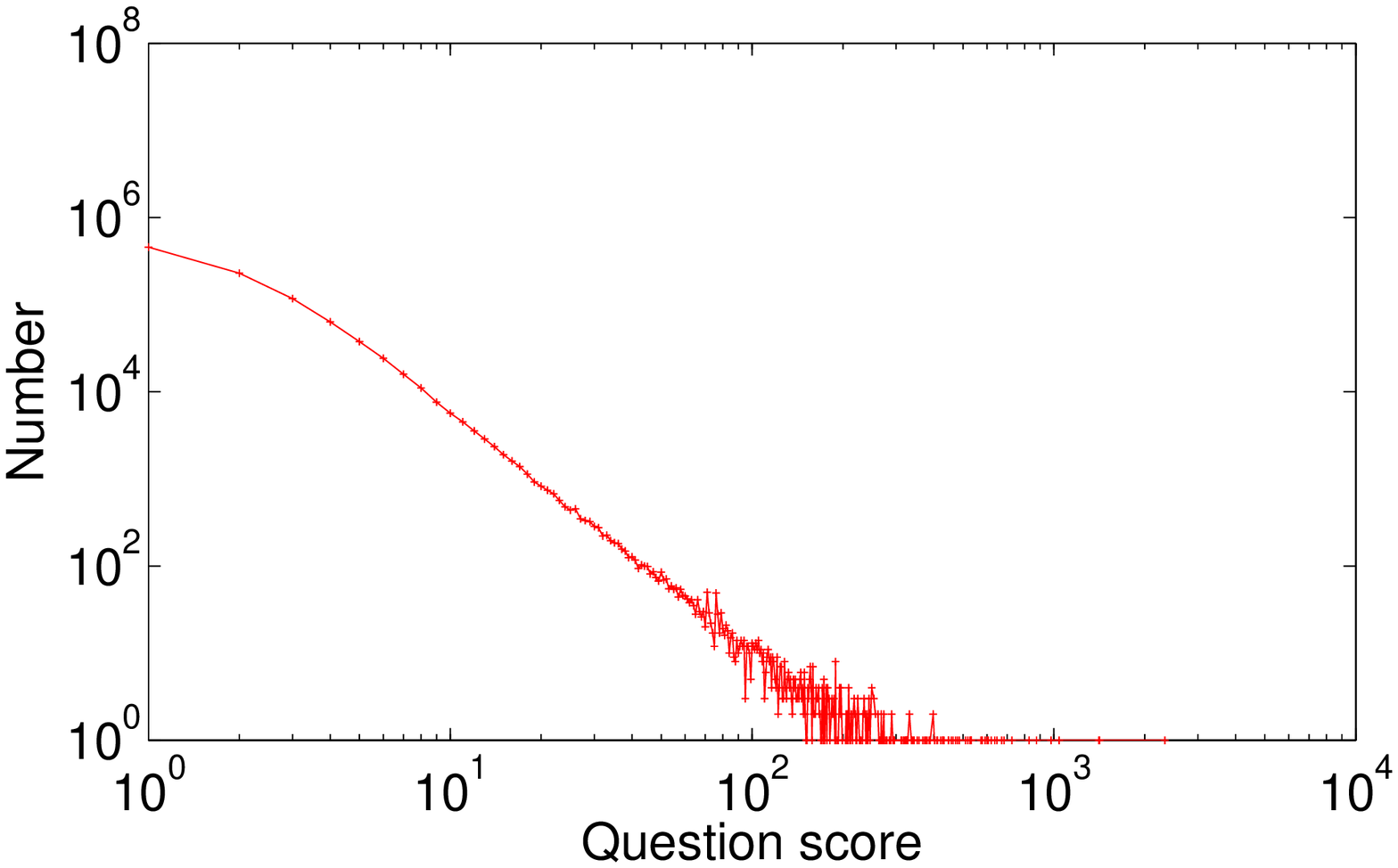}}
  \hspace{0.1in}
  \subfigure[Answers]{
    \includegraphics[width=1.5in]{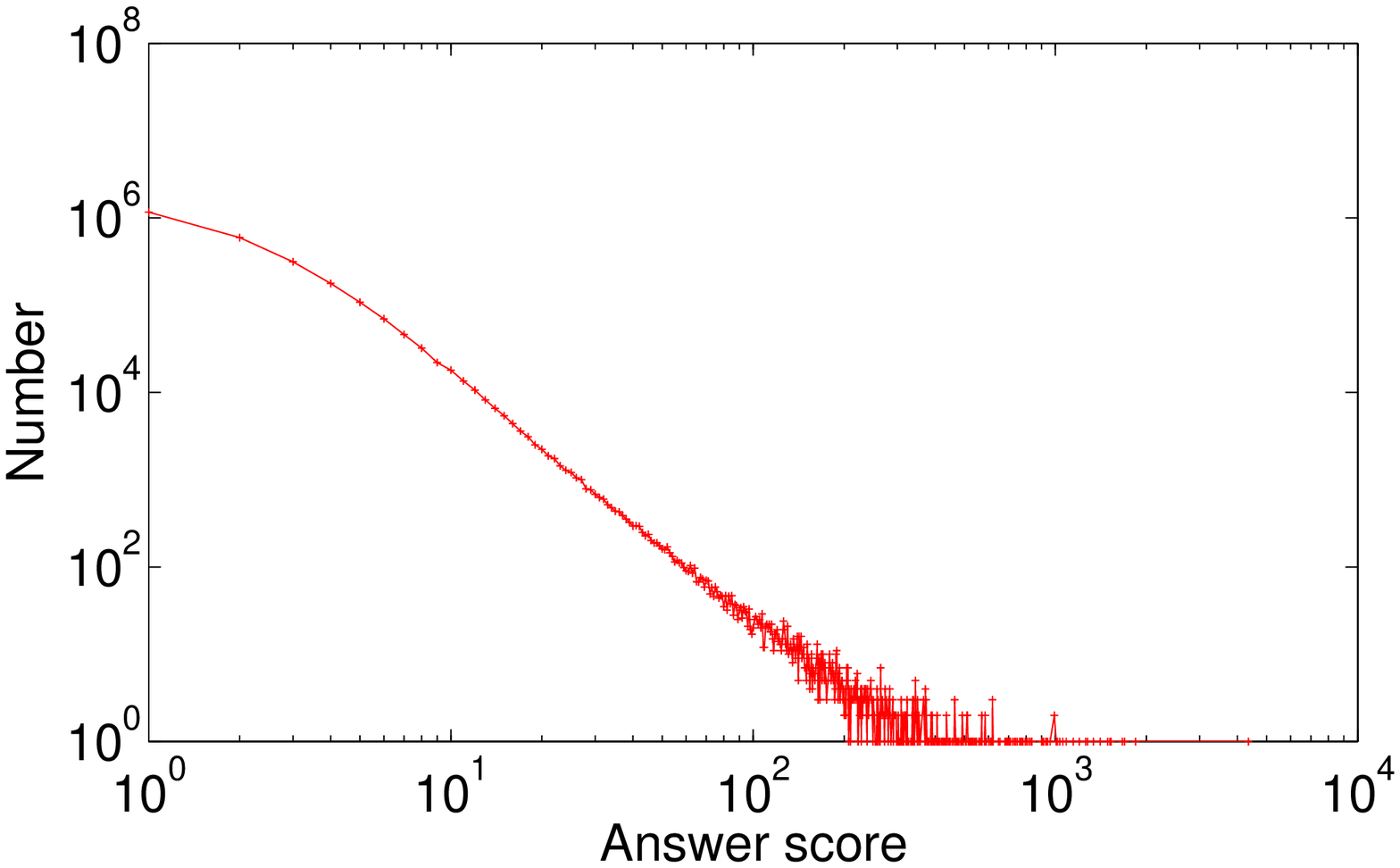}}
  \caption{The skewed, power-law like distribution of question quality and answer quality.}
\label{F:qualitydistribution}
\end{figure}

\begin{figure}[t]
  \centering
  \subfigure[Average answer quality vs. Question quality ($r = 0.4513, p < e^{-20}$)]{
    \includegraphics[width=1.5in]{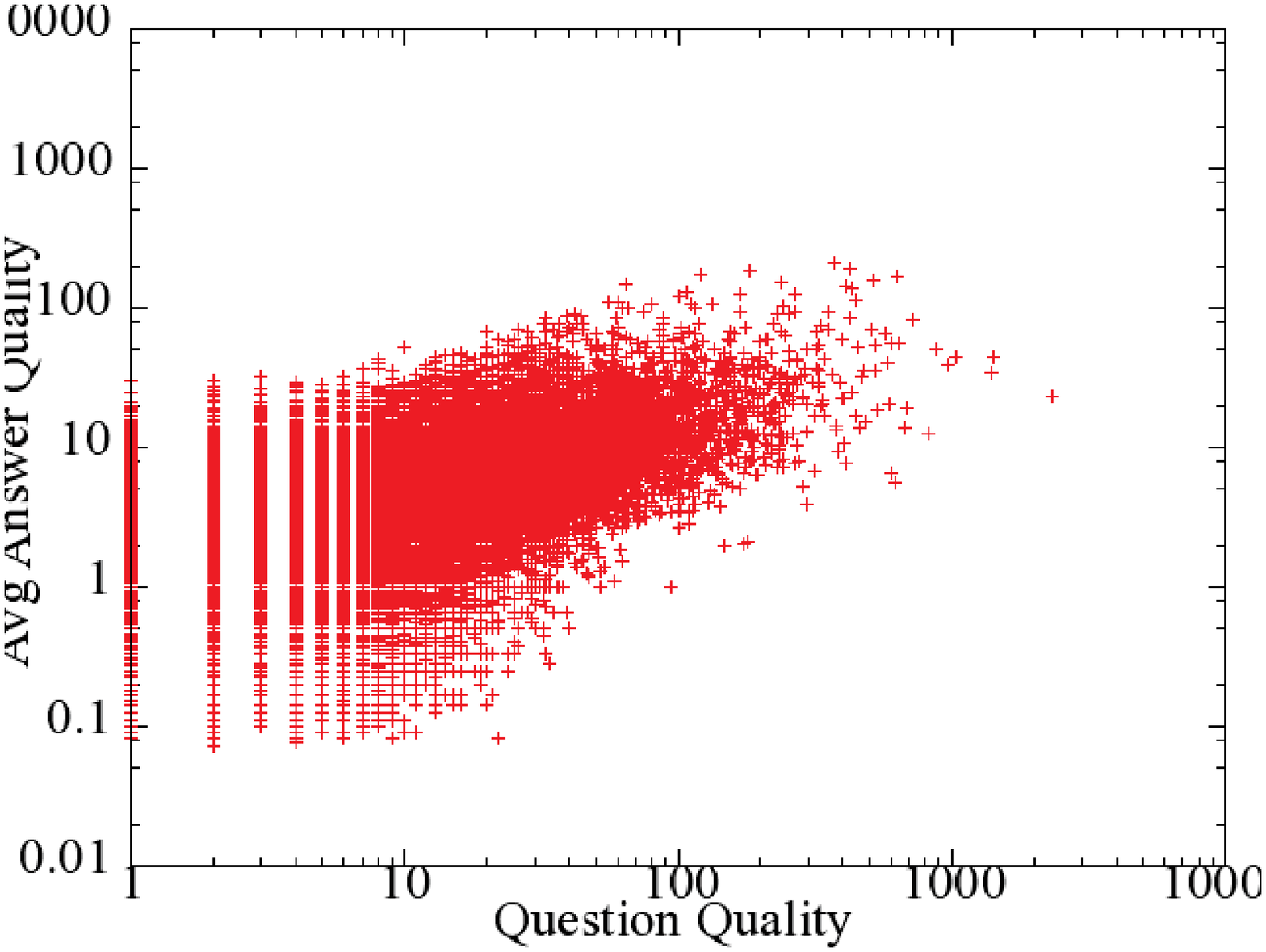}}
  \hspace{0.1in}
  \subfigure[Maximum answer quality vs. Question quality ($r = 0.7372, p < e^{-20}$)]{
    \includegraphics[width=1.5in]{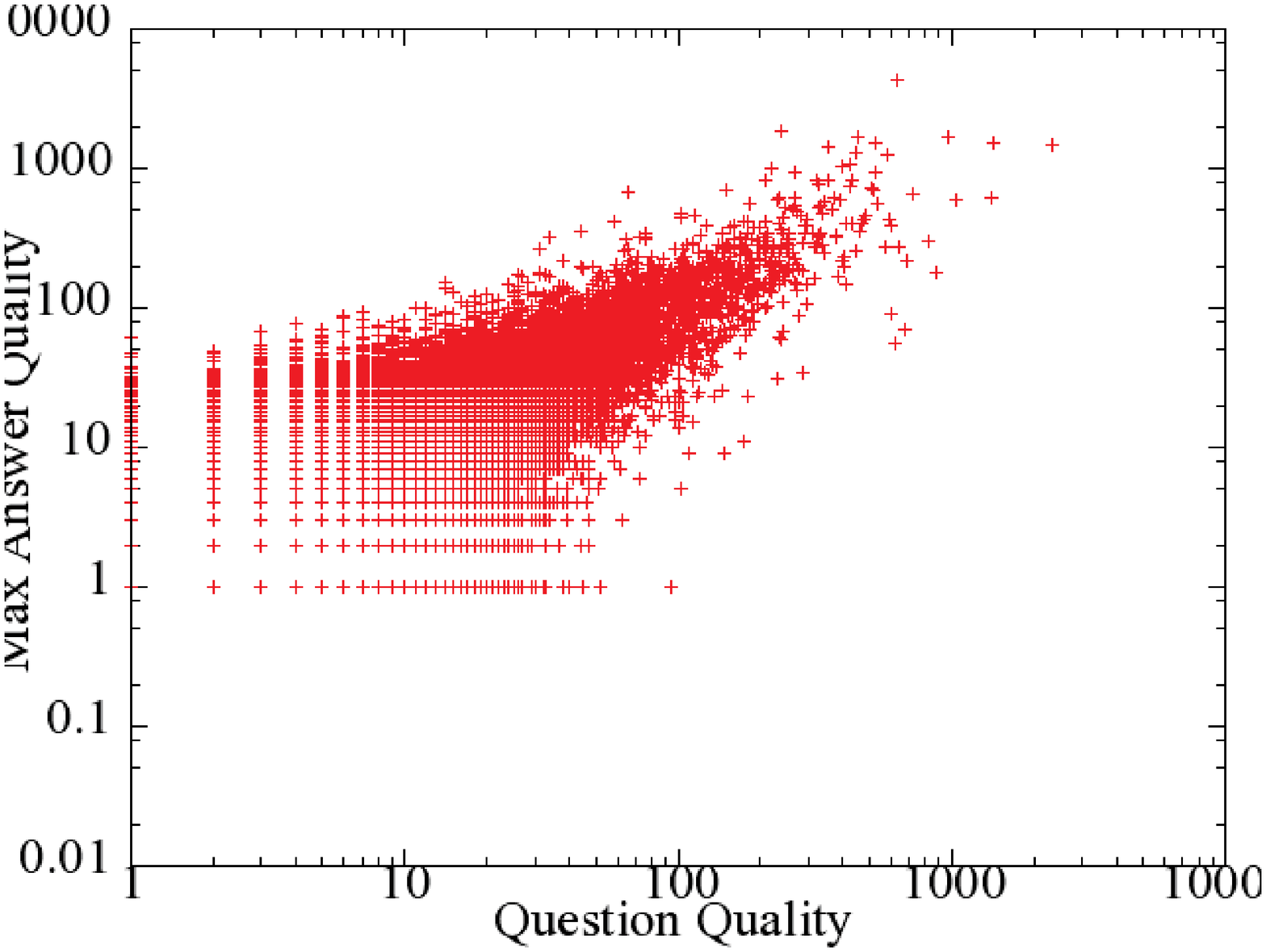}}
  \caption{The strong quality correlation between questions and their answers in Stack Overflow.}
\label{F:correlation}
\end{figure}


\begin{table*}[!t]
\caption{Quality distribution over the divided bins.}
\label{T:bins}\centering
\begin{tabular}{c||c|c|c|c|c|c|c|c|c|c|c}
  \hline
  Bins & A1 & A2 & A3 & A4 & A5 & A6 & A7 & A8 & A9 & A10 & A11 \\ \hline\hline
  Q1 & 4.58\% & 23.56\% & 23.10\% & 22.60\% & 11.03\% & 5.99\% & 3.42\% & 4.70\% & 1.03\% & 0.01\% & 0.01\% \\
  Q2 & 1.97\% & 27.71\% & 28.12\% & 23.19\% & 9.80\% & 4.47\% & 2.19\% & 2.33\% & 0.21\% & 0.00\% & 0.00\% \\
  Q3 & 1.96\% & 18.18\% & 24.64\% & 26.35\% & 13.33\% & 6.85\% & 3.64\% & 4.53\% & 0.52\% & 0.00\% & 0.00\% \\
  Q4 & 2.05\% & 13.88\% & 19.98\% & 25.86\% & 15.26\% & 8.86\% & 5.36\% & 7.56\% & 1.18\% & 0.00\% & 0.00\% \\
  Q5 & 2.13\% & 11.80\% & 17.24\% & 23.35\% & 15.40\% & 10.03\% & 6.61\% & 11.23\% & 2.21\% & 0.00\% & 0.00\% \\
  Q6 & 2.26\% & 10.66\% & 15.64\% & 21.32\% & 14.59\% & 10.21\% & 7.03\% & 14.47\% & 3.82\% & 0.00\% & 0.00\% \\
  Q7 & 2.34\% & 9.75\% & 14.78\% & 20.07\% & 13.81\% & 9.72\% & 7.32\% & 16.56\% & 5.63\% & 0.01\% & 0.00\% \\
  Q8 & 2.43\% & 8.86\% & 13.13\% & 18.32\% & 12.58\% & 9.05\% & 6.78\% & 18.30\% & 10.54\% & 0.03\% & 0.00\% \\
  Q9 & 2.51\% & 6.94\% & 10.48\% & 15.41\% & 10.83\% & 7.86\% & 5.96\% & 16.81\% & 22.06\% & 1.06\% & 0.08\% \\
  Q10 & 2.26\% & 4.86\% & 7.49\% & 12.39\% & 8.78\% & 7.10\% & 5.14\% & 15.93\% & 26.10\% & 7.00\% & 2.94\% \\
  Q11 & 2.43\% & 3.21\% & 5.39\% & 8.73\% & 6.91\% & 6.18\% & 4.05\% & 14.97\% & 28.62\% & 8.03\% & 11.49\% \\\hline
\end{tabular}
\end{table*}

Based on the above pre-processing steps, we now empirically study the post quality in Stack Overflow. We use the score of the post (which is the difference between the number of up-votes and the number of down-votes on the post) as the quality indicator, and first show the quality distribution of questions and answers in Fig.~\ref{F:qualitydistribution}. The x-axis is of log scale, and the non-positive scores are omitted from the figures. As we can see, both question quality and answer quality follow the power-law distribution. This result means that a large portion of posts in Stack Overflow receive little attention, e.g., less than 5 votes.

Next, we study the overall correlation between the quality of questions and that of their answers in Fig.~\ref{F:correlation}, where the Pearson correlation coefficient $r$ is also computed. In Fig.~\ref{F:correlation}, we consider the question quality with both the average answer quality and the maximum answer quality, since there could be multiple answers for one question. As we can see from the figures, the quality of questions and their answers are strongly correlated in both cases.


\begin{figure}[t]
  \centering
  \subfigure[Questions]{
    \includegraphics[width=1.5in]{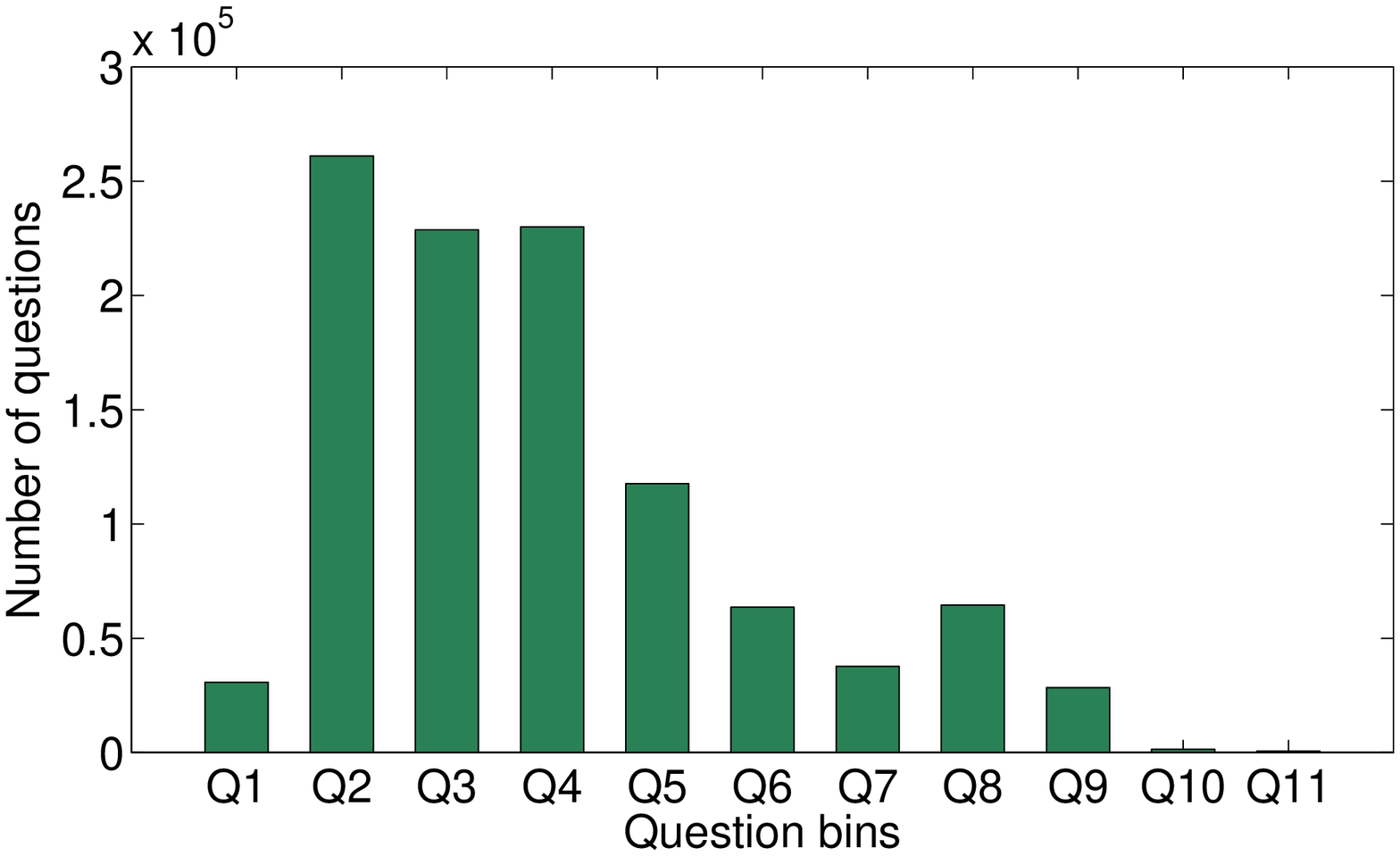}}
  \hspace{0.1in}
  \subfigure[Answers]{
    \includegraphics[width=1.5in]{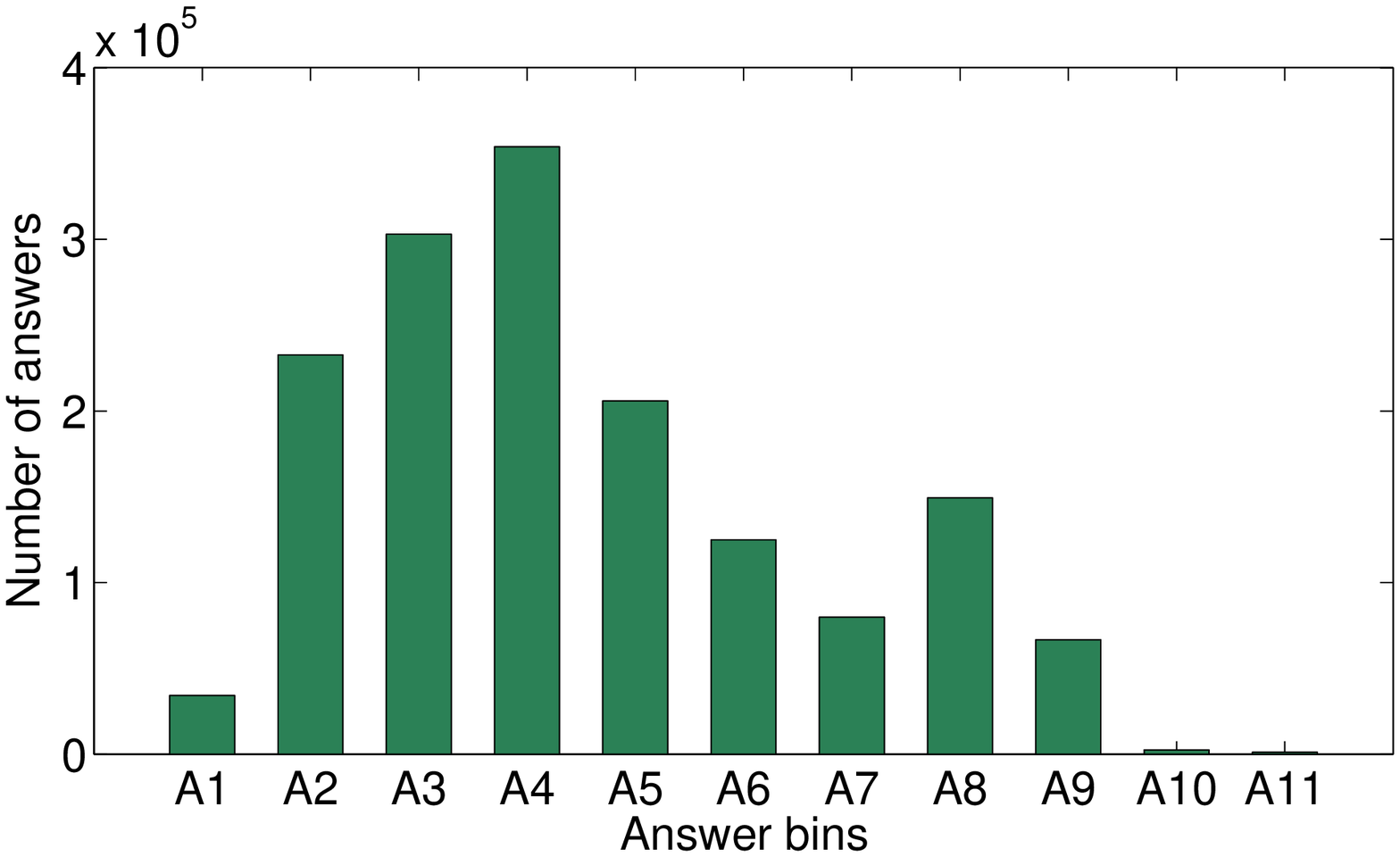}}
  \caption{Post distribution over bins.}
\label{F:bindistribution}
\end{figure}


To gain a deeper understanding of the quality correlation, we further divide questions and answers into several bins based on their scores. Specially, questions and answers are divided into 11 bins (i,e, $Q1, Q2, ..., Q11$, and $A1, A2, ..., A11$) where the score $s$ in each bin is in the rage of $s<0, s=0, s=1, s=2, s=3, s=4, s=5, 6 \leqslant s \leqslant 10, 11 \leqslant s \leqslant 50, 51 \leqslant s \leqslant 100, s>100$, respectively.

As mentioned before, a large portion of questions and answers are of low score (e.g., around 40\% question scores are 0, and around 25\% question scores are 1). Such a skewed distribution may mislead the quality prediction algorithms. For instance, if we simply predict everything as low quality, we would get a high prediction accuracy. However, such an algorithm would miss all the high-quality questions/answers. To address this issue, we take an additional pre-processing step to make the distribution more even by deleting some low-quality posts. Specially, we randomly cut two-third of the 0-score questions in $Q2$ and half of the 1-score questions in $Q3$; and then also randomly cut two-third of the 0-score answers in $A2$ and half of the 1-score answers in $A3$ as well as the suspending answers whose corresponding questions are deleted. After this step, we have 1,064,142 questions and 1,554,045 answers, and the quality distribution over bins is shown in Fig.~\ref{F:bindistribution}. As we can see, the distribution over bins is more balanced now compared to the power-law distribution. The positive quality correlation still exists in this dataset (e.g., the Pearson correlation coefficient is $r = 0.2285, p < e^{-20}$ in the average case), and we will use this dataset as our final dataset in the experiments.

Based on the divided bins, we next study the distribution of answers over each question bin. That is, for questions in bin $Qi$, we show the percentage of associated answers over the 11 answer bins in Table~\ref{T:bins}. Several interesting observations can be found from the table. First, the major questions and answers are in the diagonal, which supports the positive correlation between question quality and answer quality. Second, the table is very sparse in the top-right corner, which verifies that it is rare to have high-quality answers when the question is of low quality. Finally, the major high-quality answers are in the right-bottom corner of the table, which means their questions are also of high quality. However, there are a few very high-quality answers for the questions with negative quality scores (i.e., questions in $Q1$). This is probably because the answers successfully identify the vulnerability of the questions and thus gain support from the community members.

Overall, we conclude from Table~\ref{T:bins} as well as the above analysis that the positive quality correlation between questions and their answers strongly exists in Stack Overflow.

\section{Quality Prediction: Problem Definitions}

\begin{table}[!h]
\caption{Symbols.}
\label{T:symbols}\centering
\begin{tabular}{|l|l|}
  \hline
  Symbol & Definition and Description \\ \hline
  $\mat{X}_q, \mat{X}_a$ & the feature matrix for questions and answers \\
  $\mat{X}_{\tilde{q}}, \mat{X}_{\tilde{a}}$ & the feature matrix with transferred features \\
  $\textbf{y}_q, \textbf{y}_a$ & the real quality of questions and answers \\
  $\boldsymbol{\beta}_q, \boldsymbol{\beta}_a$ & the coefficients for the features \\
  $\mat{M}$ & the association matrix of questions and answers \\
  $\tilde{\mat{M}}$ & the row-normalized matrix of \textbf{M} \\
  $\mat{M}'$ & the transpose of matrix $\textbf{M}$ \\
  $\mat{M}(i,j)$ & the element at the $i^{th}$ row and $j^{th}$ column of $\textbf{M}$ \\
  $\mat{M}(i,:)$ & the $i^{th}$ row of matrix $\textbf{M}$ \\ \hline
  $n_q, n_a$ & the number of questions and answers \\
  $m_1, m_2$ & the maximum iteration number \\
  $\xi_1, \xi_2$ & the threshold to terminate the iteration \\ \hline
\end{tabular}
\end{table}

In this section, we present the problem definition for quality prediction of questions/answers in CQA websites. Before that, let us first introduce the notations that would be used throughout the paper.
The notations are listed in Table~\ref{T:symbols}. Following conventions, we use bold capital letters for matrices, and bold lower case letters for vectors. For example, we use $\mat{X}_q$ to denote the feature matrix for questions where each line contains the feature vector for the corresponding question. We use the $n_q \times n_a$ matrix $\mat{M}$ to denote the association matrix of questions and answers where non-zero element $\mat{M}(i,j)$ indicates that the $j^{th}$ answer belongs to the $i^{th}$ question. Similar to Matlab, we also denote the $i^{th}$ row of matrix $\mat{M}$ as $\mat{M}(i,:)$, and the transpose of a matrix with a prime.

Based on the above notations, the prediction problem for questions could be defined as follows:
\begin{problem}{The Question Quality Prediction}\label{P:basic}
\begin{description}
    \item[Given:] the question feature matrix $\mat{X}_q$ where each line represents the feature vector for the corresponding question, and the vector of question quality scores $\mat{y}_q$;
    \item [Output:] the quality of a new question.
\end{description}
\end{problem}

Analogous definition for answer quality prediction can be defined, and we omit the definition for brevity. As mentioned in the previous section, we consider the answers as well as the available information as soon as the question is posted. In this work, we fix the time window to 24 hours, and therefore the feature matrices $\mat{X}_q$ and $\mat{X}_a$ should only contain the information that is available in the first 24 hours after the question is posted.

One of our key findings in the previous section is the strong positive quality correlation between questions and their answers. As a result, we take the association matrix $\mat{M}$ into account, and predict the question and answer quality jointly. Based on the problem definition above, we now define our quality co-prediction problem as follows:
\begin{problem}{The Question and Answer Quality Co-Prediction}\label{P:final}
\begin{description}
    \item[Given:] the question feature matrix $\mat{X}_q$, the answer feature matrix $\mat{X}_a$, the vector of question quality scores $\mat{y}_q$, the vector of answer quality scores $\mat{y}_a$, and the association matrix $\mat{M}$;
    \item [Output:] the quality of a new question and its answers.
\end{description}
\end{problem}

As we can see from the above problem definition, compared to the separate prediction problem, the only additional input of Problem~\ref{P:final} is the association matrix $\mat{M}$. We will present the methods to solve Problem~\ref{P:basic} and Problem~\ref{P:final} in the next section.

\section{Quality Prediction: The Proposed Approach}
In this section, we present our algorithms to solve the problems defined in the previous section. We start with the baseline algorithms which predict the question quality and answer quality separately. Then, we present our co-prediction methods (\cop) to jointly predict the question and answer quality, for both the regression case (e.g., to infer numerical quality scores) and the classification case (e.g., to differentiate the high-quality questions/answers from those of low quality).

\subsection{Preliminaries: Separate Methods}

Here, we first introduce the separate method which could be used for the quality prediction in Problem~\ref{P:basic}. Specially, the problem can be formulated as the following optimization problem:
\begin{eqnarray}\label{E:se-formulation}
\boldsymbol\beta_q = \textrm{argmin}_{\boldsymbol\beta_q} \sum_{i=1}^{n_q} g(\mat{X}_q(i,:) \boldsymbol\beta_q, \mat{y}_q(i)) + \lambda ||\boldsymbol\beta_q||^2
\end{eqnarray}
where $g$ indicates the loss function. We also add a regularization term $||\boldsymbol\beta_a||^2$ which is controlled by $\lambda$ to avoid over-fitting.

In the above formulation, various loss functions could be used. For example, for the purpose of classification, we may choose logistic function for $g$; for regression, we can choose the square loss, i.e., $g_{square}(x,y) = (x-y)^2$. Notice that in the latter case, the formulation becomes the standard ridge regression problem and it can be solved by the following closed-form solution:
\begin{eqnarray}\label{E:se-solution}
\boldsymbol{\beta}_q = (\mat{X}_q' \mat{X}_q + \lambda\mat{I})^{-1} \mat{X}_q' \mat{y}_q
\end{eqnarray}

Analogous solution for answer quality prediction can be similarly written. We denote such method as {\em Separate} method, and will compare it with our co-prediction methods in the experiments.

\subsection{Co-Prediction: Basic Strategies}




Next, we discuss some basic strategies that we explore to leverage the observed quality correlation, before we present our \cop\ algorithms. There are three basic strategies behind our \cop\ algorithms. We first summarize them together with the intuitions behind each strategy as follows:
\begin{itemize}
\item \textbf{S1:} The first strategy considers the feature space. Because the quality of questions and their answers is correlated, the features for question prediction are potentially useful for answer prediction. As a result, we transfer the question features to $\mat{M}' \mat{X}_q$ and add these features for answer quality prediction. Namely, we use $\mat{X}_{\tilde{a}} = [\mat{X}_a, \mat{M}' \mat{X}_q]$ to represent the new feature matrix for answers. Similarly, we transfer the answer features to $\tilde{\mat{M}} \mat{X}_a$ and incorporate these features with $\mat{X}_q$ as $\mat{X}_{\tilde{q}} = [\mat{X}_q, \tilde{\mat{M}} \mat{X}_a]$. Notice that we use the row-normalized $\tilde{\mat{M}}$ matrix in the latter case. In other words, for a question with multiple answers, we take the average of the features from these answers.
\item \textbf{S2:} The second strategy takes into account the link between the feature space and the label space. That is, we could iteratively use the estimated question score $\mat{\hat y}_q$ as a feature for answer prediction, and the estimated answer score $\mat{\hat y}_a$ as a feature for question prediction. The intuition behind this strategy is that if the estimated quality score $\mat{\hat y}_q$ is close to the real quality score $\mat{y}_q$, then $\mat{\hat y}_q$ would also be high-correlated with $\mat{y}_a$.
\item \textbf{S3:} The third strategy considers the label space. For a pair of question and answer, we could directly maximize the quality correlation or minimize the quality difference between them. In this work, we try to minimize the difference between the predicted score of a question and that of its answer. Namely, we require that $\mat{\hat y}_q \approx \tilde{\mat{M}} \mat{\hat y}_a$, where we also constrain that the question score is close to the average score of its answers. We study the maximum answer score as well, and similar results are observed. For brevity, we will focus on the average case in this work.
\end{itemize}

With these three strategies, we will focus on the regression problem (i.e., to infer numerical quality scores) and the classification problem (i.e., to differentiate the high quality questions/answers from those of low quality) in the next two subsections, respectively.

\subsection{Proposed Approach for Regression}

\begin{algorithm}[t]
\caption{\copit\ algorithm.}\label{A:iterative}
\begin{algorithmic}[1]
  \REQUIRE {$\mat{X}_{q}$, $\mat{X}_{a}$, $\mat{y}_q$, $\mat{y}_a$, and $\mat{M}$}
  \ENSURE {$\boldsymbol{\beta}_q$ and $\boldsymbol{\beta}_a$}
  \STATE $\mat{X}_{\tilde{q}} \leftarrow [\mat{X}_q, \tilde{\mat{M}} \mat{X}_a]$;
  \STATE $\mat{X}_{\tilde{a}} \leftarrow [\mat{X}_a, \mat{M}' \mat{X}_q]$;
  \STATE $\boldsymbol{\beta}_a$ $\leftarrow$ ($\mat{y}_a, \mat{X}_{\tilde{a}}$) by Eq.~\eqref{E:se-formulation};
  \STATE $\mat{\hat y}_a$ $\leftarrow$ $\mat{X}_{\tilde{a}} \boldsymbol{\beta}_a$;
  \WHILE {not convergent}
    \STATE $\boldsymbol{\beta}_q$ $\leftarrow$($\mat{y}_q, [\mat{X}_{\tilde{q}}, \tilde{\mat{M}}\mat{\hat y}_a]$) by Eq.~\eqref{E:se-formulation};
    \STATE $\mat{\hat y}_q$ $\leftarrow$ $[\mat{X}_{\tilde{q}}, \tilde{\mat{M}}\mat{\hat y}_a] \boldsymbol{\beta}_q$;
    \STATE $\boldsymbol{\beta}_a$ $\leftarrow$ ($\mat{y}_a, [\mat{X}_{\tilde{a}}, \mat{M}'\mat{\hat y}_q]$) by Eq.~\eqref{E:se-formulation};
    \STATE $\mat{\hat y}_a$ $\leftarrow$ $[\mat{X}_{\tilde{a}}, \mat{M}'\mat{\hat y}_q] \boldsymbol{\beta}_a$;
  \ENDWHILE
  \RETURN $\boldsymbol{\beta}_q$ and $\boldsymbol{\beta}_a$;
\end{algorithmic}
\end{algorithm}

Here, we consider the continuous case of the co-prediction problem where we want to infer numerical scores for each question and their associated answers. We propose \copit\ which is based on {\bf S1} and {\bf S2}. The detail of the proposed \copit\ is shown in Alg.~\ref{A:iterative}. As we can see in the algorithm, \copit\ first incorporates the transferred features, estimates the answer quality $\mat{\hat y}_a$, and then starts iteration. In each iteration, \copit\ alternatively uses $\mat{\hat y}_a$ and $\mat{\hat y}_q$ as features for the question quality prediction and answer quality prediction, respectively. We will stop the iteration when the $L_2$ norm between successive estimates of both $\boldsymbol\beta_q$ and $\boldsymbol\beta_a$ is below our threshold $\xi_1$, or the maximum iteration number $m_1$ is achieved.


\subsection{Proposed Approach for Classification}

Here, we consider the binary case of the co-prediction problem where want to differentiate the high-quality questions/answers from those of low quality. We first show the general procedure for solving the co-prediction problem followed by a special case, and then briefly analyze the effectiveness and efficiency of the algorithms.

\subsubsection{The General Procedure}

The proposed approach is based on {\bf S1} and {\bf S3}. To be specific, we propose a new optimization formulation for question and answer quality co-prediction problem. That is, after transferring the features, we add the $\mat{\hat y}_q \approx \tilde{\mat{M}} \mat{\hat y}_a$ constraint as an additional term into the optimization formulation:
\begin{eqnarray}\label{E:co-formulation}
\mathcal{L} &=& \min_{\boldsymbol\beta_q, \boldsymbol\beta_a} \sum_{i=1}^{n_q} g(\mat{X}_{\tilde{q}}(i,:) \boldsymbol\beta_q, \mat{y}_q(i)) + \sum_{i=1}^{n_a} g(\mat{X}_{\tilde{a}}(i,:) \boldsymbol\beta_a, \mat{y}_a(i)) \nonumber\\
 & & + \eta \sum_{i=1}^{n_q} h(\mat{X}_{\tilde{q}}(i,:) \boldsymbol\beta_q, \tilde{\mat{M}}(i,:) \mat{X}_{\tilde{a}} \boldsymbol\beta_a) + \lambda (||\boldsymbol\beta_q||^2_2 + ||\boldsymbol\beta_a||^2_2)
\end{eqnarray}
where $h$ indicates the loss function of the additional quality correlation term, and $\eta$ is parameter to control the importance of this term.
Compared to \copit, an advantage of the above formulation is that it can deal with the sparsity problem when some of the quality scores are not available. For example, in the extreme case when there are no answer scores available, Eq.~\eqref{E:co-formulation} can still predict the scores of new answers due to the constraint of the quality correlation term. We will experimentally evaluate this in section~\ref{sec:experiments}.

Next, let us further discuss the loss functions $g$ and $h$ in Eq.\eqref{E:co-formulation}. In addition to the square loss we mentioned before, other loss functions could be applied under our classification setting. The intuition is that while the square loss tends to minimize the difference between the real quality score and the estimated quality score, the consistency between the real quality label and the estimated label is also important for the classification task. Therefore, we divide the posts into two classes: high quality with label +1 and low quality with label -1, and then consider the sigmoid loss function:
\begin{eqnarray}\label{E:sigmoid}
g_{sigmoid}(x,y) = 1/(1 + \exp(xy)).
\end{eqnarray}
By setting $g$ and $h$ as square loss and sigmoid loss, we could derive four variants from Eq.~\eqref{E:co-formulation} as shown in Table~\ref{T:combinations}.

\begin{table}[!h]
\caption{The four variants derived from Eq.~\eqref{E:co-formulation}.}
\label{T:combinations}\centering
\begin{tabular}{c||c|c}
  \hline
  Denoter & $g$ & $h$ \\ \hline
  \copqq & square loss & square loss \\
  \copqg & square loss & sigmoid loss \\
  \copgg & sigmoid loss & sigmoid loss \\
  \copgq & sigmoid loss & square loss \\\hline
\end{tabular}
\end{table}

\begin{algorithm}[t]
\caption{Algorithm skeleton for \copqq, \copqg, \copgg, and \copgq\ (See the appendix for the details).}\label{A:co-gd}
\begin{algorithmic}[1]
  \REQUIRE {$\mat{X}_{q}$, $\mat{X}_{a}$, $\mat{y}_q$, $\mat{y}_a$, and $\mat{M}$}
  \ENSURE {$\boldsymbol{\beta}_q$ and $\boldsymbol{\beta}_a$}
  \STATE $\mat{X}_{\tilde{q}} \leftarrow [\mat{X}_q, \tilde{\mat{M}} \mat{X}_a]$;
  \STATE $\mat{X}_{\tilde{a}} \leftarrow [\mat{X}_a, \mat{M}' \mat{X}_q]$;
  \STATE $\boldsymbol{\beta}_q$ $\leftarrow$ ($\mat{y}_q, \mat{X}_{\tilde{q}}$) by Eq.~\eqref{E:se-formulation};
  \STATE $\boldsymbol{\beta}_a$ $\leftarrow$ ($\mat{y}_a, \mat{X}_{\tilde{a}}$) by Eq.~\eqref{E:se-formulation};
  \WHILE {not convergent}
    \STATE $\boldsymbol\beta_q$ $\leftarrow$ $\boldsymbol\beta_q - \gamma \frac{\partial{\mathcal{L}}}{\partial{\boldsymbol\beta_{q}}}$;
    \STATE $\boldsymbol\beta_a$ $\leftarrow$ $\boldsymbol\beta_a - \gamma  \frac{\partial{\mathcal{L}}}{\partial{\boldsymbol\beta_{a}}}$;
  \ENDWHILE
  \RETURN $\boldsymbol{\beta}_q$ and $\boldsymbol{\beta}_a$;
\end{algorithmic}
\end{algorithm}

Finally, Alg.~\ref{A:co-gd} shows the skeleton of the general algorithms to solve the four variants in Table~\ref{T:combinations}. The details of step 6-7 with various loss functions are presented in the appendix for completeness. As we can see from the algorithm skeleton, after initializing $\boldsymbol{\beta}_q$ and $\boldsymbol{\beta}_a$, we adopt the batch gradient descent method to iteratively update the coefficients.
We will stop the iteration when the $L_2$ norm between successive estimates of both $\boldsymbol\beta_q$ and $\boldsymbol\beta_a$ is below our threshold $\xi_2$, or the maximum iteration number $m_2$ is achieved.


\subsubsection{A Special Case}

The optimization problem in Eq.~\eqref{E:co-formulation} is non-convex in general. However, if we set both $g$ and $h$ as square loss, the optimization problem becomes convex and we can have the closed-form solution for \copqq\ as follows:
\begin{eqnarray}\label{E:co-solution}
{\begin{pmatrix}
            \boldsymbol\beta_{q} \\
            \boldsymbol\beta_{a} \\
  \end{pmatrix}} =
{\begin{pmatrix}
            (\eta + 1) \mat{X}_{\tilde{q}}' \mat{X}_{\tilde{q}} + \lambda \mat{I} & -\eta \mat{X}_{\tilde{q}}' \tilde{\mat{M}} \mat{X}_{\tilde{a}} \\
            -\eta \mat{X}_{\tilde{a}}' \tilde{\mat{M}}' \mat{X}_{\tilde{q}} & \mat{X}_{\tilde{a}}' \mat{X}_{\tilde{a}} + \eta \mat{X}_{\tilde{a}}' \tilde{\mat{M}}' \tilde{\mat{M}} \mat{X}_{\tilde{a}} + \lambda \mat{I} \\
   \end{pmatrix}^{-1}}
{\begin{pmatrix}
            \mat{X}_{\tilde{q}}' \mat{y}_{q} \\
            \mat{X}_{\tilde{a}}' \mat{y}_{a} \\
   \end{pmatrix}}
\end{eqnarray}


\subsubsection{Algorithm Analysis}

Here, we briefly analyze the effectiveness and efficiency of our algorithms for the optimization problem in Eq.~\eqref{E:co-formulation}.

The effectiveness of the proposed algorithms can be summarized in Lemma~\ref{L:effectiveness}, which says that the algorithms in Alg.~\ref{A:co-gd} can find a local minima solution, and Eq.~\eqref{E:co-solution} can find the global minima solution.

\begin{lemma}{\bf {Effectiveness of \cop}}.\label{L:effectiveness}
The \cop\ algorithms in Alg.~\ref{A:co-gd} find local minima for the optimization problem in Eq.~\eqref{E:co-formulation}. In the special case, the algorithm in Eq.~\eqref{E:co-solution} finds the global minima.
\end{lemma}
\proof Omitted for brevity. \QED

The time complexity of the proposed \cop\ is summarized in Lemma~\ref{L:efficiency}, which says that \cop\ scales linearly wrt the sum of questions and answers.

\begin{lemma}{\bf {Efficiency of \cop}}.\label{L:efficiency}
By fixing the feature number in $\mat{X}_{q}$ and $\mat{X}_{a}$, and storing matrix $\tilde{\mat{M}}$ in sparse matrix format, the \cop\ algorithms in Alg.~\ref{A:co-gd} require $O(n_q m_2 + n_a m_2)$ time where $m_2$ is the maximum iteration number, and the algorithm in Eq.~\eqref{E:co-solution} requires $O(n_q + n_a)$ time.
\end{lemma}
\proof Omitted for brevity. \QED

\section{Experiments}\label{sec:experiments}
In this section, we present the experimental evaluations. All the experiments are designed to answer the following questions:
\begin{itemize}
\item {\em Effectiveness}: How accurate are the proposed methods for question and answer quality prediction?
\item {\em Efficiency}: How do the proposed methods scale?
\end{itemize}

\subsection{Experiment Setup}

We first describe the features we use for our question quality prediction and answer quality prediction, and the selected features are shown in Table~\ref{T:qfeature} and~\ref{T:afeature}, respectively. Notice that our goal is not to find the most useful features for separate question/answer quality prediction, but to show that the quality correlation between questions and their answers could really improve the prediction accuracy.
Therefore, as we can see from the tables, we choose some commonly used features in Q\&A quality prediction. For example, questioners' reputation could reflect their ability of posting a high-quality question, questioners could be more professional as they post more questions, etc. The number of comments and the length of the post are also widely used by existing methods. Notice that we do not include the vote information in selected features because the quality score is defined as the difference between up-votes and down-votes in Stack Overflow.

\begin{table}[!t]
\caption{Selected features for question quality prediction.}
\label{T:qfeature}\centering
\begin{tabular}{l}
  \hline
  Feature Description \\ \hline
  Questioner's reputation when the question is posted \\
  \# of Questioner's previous questions when the question is posted \\
  \# of answers received in 24 hours after the question is posted \\
  \# of favorites received in 24 hours after the question is posted \\
  \# of comments received in 24 hours after the question is posted \\
  The length of the question \\
  The length of the title \\\hline
\end{tabular}
\end{table}

\begin{table}[!t]
\caption{Selected features for answer quality prediction.}
\label{T:afeature}\centering
\begin{tabular}{l}
  \hline
  Feature Description \\ \hline
  Answerer's reputation when the answer is posted \\
  \# of Answerer's previous answers when the answer is posted \\
  \# of comments received in 24 hours after its question is posted \\
  The length of the answer \\\hline
\end{tabular}
\end{table}

Now, we already have the input matrices $\mat{X}_q$ and $\mat{X}_a$ based on the selected features, and the next input is the quality score. Here, in the regression setting (i.e., to infer continuous numerical quality scores), we normalize the quality scores between 0 and 1. In the classification setting (i.e., to predict the binary labels for high-quality and low-quality questions/answers), we classify the posts whose quality score is equal to or smaller than 0 as low-quality posts, and those whose quality score is equal to or larger than 5 as high-quality posts.

In terms of evaluation metrics, we adopt the root mean square error (RMSE) and the prediction error between the real quality and the estimated quality in the continuous case and the binary case, respectively. Specially, we will evaluate effectiveness of \copit\ in the regression setting, and study the four variants in Table~\ref{T:combinations} (i.e., \copqq, \copqg,\copgg, and \copgq) in the classification setting. Unless otherwise stated, for all the results reported here, we fix $m_1 = m_2 = 20$, $\xi_1 = \xi_2 = 10^{-9}$, and $\gamma = 10^{-6}$.

In addition to the {\em Separate} method we outline in section 4.1, we are not aware of any existing work for joint prediction of question and answer quality, with the only exception of the {\em CQA-MR} method~\cite{bian2009learning}. {\em CQA-MR} aims to simultaneously classify the user reputation, and the quality of answers and questions in a mutually reinforced manner. We would like to point out that although both {\em CQA-MR} and our \cop\ aim to improve classification accuracy by {\em co-prediction}, there are two important differences. First, while {\em CQA-MR} implements the {\em co-prediction} through the label space (by propagating the labels through user-question-answer graph), our \cop\ does so through both label space and feature space, as well as the link between them (see {\bf S1-S3} in Section 4.2). Second, our \cop\ finds either a local or a global minima for Eq.~\eqref{E:co-formulation} (depending on the specific loss functions). In contrast, {\em CQA-MR} alternates between propagating label and maximizing the corresponding conditional likelihood. Consequently, it is not clear what overall cost function {\em CQA-MR} aims to optimize and whether or not the overall procedure converges. As we will show below, such subtle differences in our \cop\ lead to a significant performance improvement.

\subsection{Effectiveness Results}

For effectiveness, we choose $K\%$ questions and their associated answers as the training set, and use the rest as the test set. All the effectiveness results reported are the average of 10 experiments.


\begin{figure}[t]
  \centering
    \includegraphics[width=2.5in]{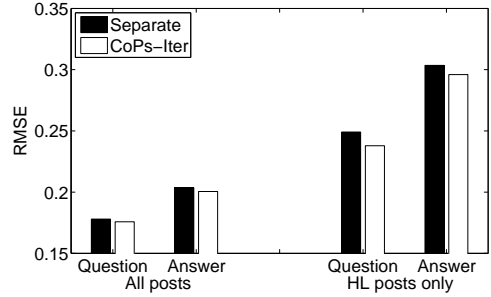}
  \caption{The RMSE results of \copit\ in the regression setting. \copit\ outperforms the {\em Separate} method.}
\label{F:effectiveness-it}
\end{figure}

{\em (A) The Effectiveness of \cop\ in Regression Setting}. We first compare the RMSE results of \copit\ with the {\em Separate} method. We set $K=1$, and show the results in Fig.~\ref{F:effectiveness-it}. As we can see from the left part of the figure, although the proposed \copit\ method is better wrt RMSE, the improvement is very limited. This is probably due to the fact that still a large amount of the normalized quality scores are in the range of $[0.1, 0.5]$. As a result, we further apply the \copit\ method on the low-quality posts and high-quality posts only, and the results are also shown in Fig.~\ref{F:efficiency} (denoted as `HL posts only'). In this case, \copit\ improves the {\em Separate} method by 4.50\% in question prediction and 2.47\% in answer prediction.

\begin{table}[!t]
\caption{The prediction error results of \copqq\, \copqg, \copgg\, and \copgq\ in the classification setting. Our methods outperform the {\em Separate} method and the {\em CQA-MR} method.}
\label{T:effectiveness-classify}\centering
\begin{tabular}{c||c|c}
  \hline
   & Questions & Answers \\ \hline
  {\em Separate} & 0.2192 & 0.3396 \\
  {\em CQA-MR} & 0.2360 & 0.3416 \\
  \copqq\ & 0.2048 & 0.3010 \\
  \copqg\ & 0.2101 & 0.2941 \\
  \copgg\ & 0.2111 & 0.2939 \\
  \copgq\ & 0.2029 & 0.2950 \\\hline
\end{tabular}
\end{table}

{\em (B) The Effectiveness of \cop\ in Classification Setting}. In the classification experiment, we first compare our four variants derived from Eq.~\eqref{E:co-formulation} with the {\em Separate} method and the {\em CQA-MR} method. We still use only 1\% questions and their associated answers as the training set, and show the results in Table~\ref{T:effectiveness-classify}. As we can see, all our four methods outperform the compared methods. For example, \copgq\ improves the {\em Separate} method by 7.44\% and 13.13\% wrt prediction error of questions and answers, respectively.

\begin{figure}[t]
  \centering
  \subfigure[Question quality prediction]{
  \label{F:QAfixK-q}\centering
    \includegraphics[width=2.5in]{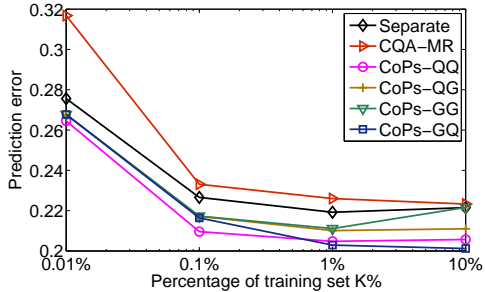}}
  \hspace{0.1in}
  \subfigure[Answer quality prediction]{
  \label{F:QAfixK-a}\centering
    \includegraphics[width=2.5in]{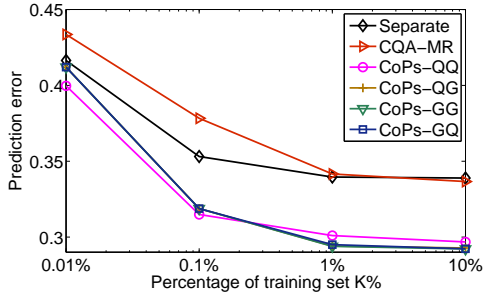}}
  \caption{The prediction error results when we vary the percentage of questions and answers in the training set. Our methods are better than the {\em Separate} method and the {\em CQA-MR} method in most cases.}
\label{F:QAfixK}
\end{figure}

Next, we vary the percentage of questions and answers in the training set, and show the prediction error over $K\%$ in Fig.~\ref{F:QAfixK}. As we can see, overall, our four methods are still better than the compared methods.  In addition, even when the available labels are very sparse (i.e., only 0.01\% questions and their associated answers are used as the training set), our methods can still predict the quality of questions/answers accurately. As to the comparison among the four methods we proposed, the \copqq\ and \copgq\ methods are a little better than the other two methods for the question quality prediction. For answer quality prediction, the four methods are all close to each other wrt prediction error.

\begin{table}[!t]
\caption{Performance gain wrt prediction error of the \copqq\ method. Both the transferred features and the quality correlation term in Eq.~\eqref{E:co-formulation} help to lower prediction error.}
\label{T:effectiveness-improvegain}\centering
\begin{tabular}{c||c|c}
  \hline
   & Questions & Answers  \\ \hline
  {\em Separate} & 0.2266 & 0.3531 \\
  {\em Separate} + transferred features & 0.2172 & 0.3188 \\
  \copqq\ & 0.2096 & 0.3149 \\\hline
\end{tabular}
\end{table}

Next, we take our \copqq\ method as an example, and show where the performance gain comes from in Table~\ref{T:effectiveness-improvegain}. Here, we fix $K=0.1$. As we can see from the table, both the transferred features and the quality correlation term in Eq.~\eqref{E:co-formulation} help to lower the prediction error. For example, in question quality prediction, while the transferred features help to lower the prediction error by 4.15\%, the quality correlation term further lowers the error by 3.50\%.

\begin{figure}[t]
  \centering
  \subfigure[Answer quality prediction]{
  \label{F:Q10deleteA-a}\centering
    \includegraphics[width=2.5in]{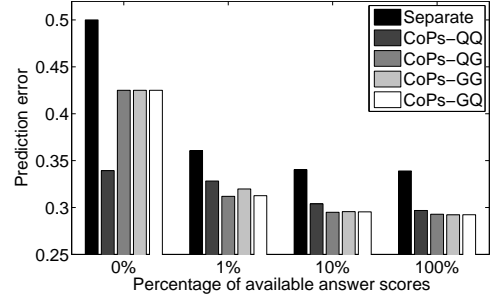}}
  \hspace{0.1in}
  \subfigure[Question quality prediction]{
  \label{F:Q10deleteA-q}\centering
    \includegraphics[width=2.5in]{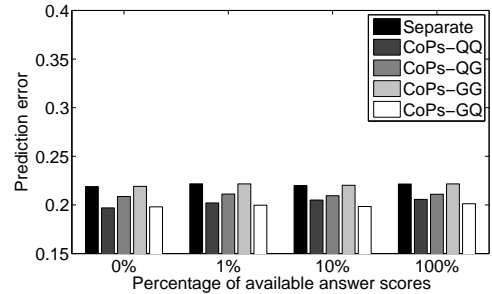}}
  \caption{The prediction error when we fix 10\% questions and their answers as the training set, and then delete available answer scores. Our methods perform better than the {\em Separate} method.}
\label{F:Q10deleteA}
\end{figure}

\begin{figure}[t]
  \centering
  \subfigure[Question quality prediction]{
  \label{F:Q10deleteQ-q}\centering
    \includegraphics[width=2.5in]{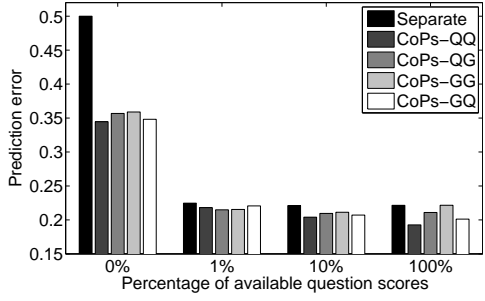}}
  \hspace{0.1in}
  \subfigure[Answer quality prediction]{
  \label{F:Q10deleteQ-a}\centering
    \includegraphics[width=2.5in]{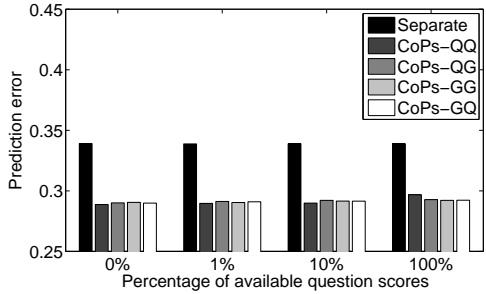}}
  \caption{The prediction error when we fix 10\% questions and their answers as the training set, and then delete available question scores. Again, our methods perform better than the {\em Separate} method.}
\label{F:Q10deleteQ}
\end{figure}

{\em (C) The Effectiveness of \cop\ to Deal With Sparsity}. As mentioned before, our methods could benefit the case when the available quality scores are sparse. We have already shown that our method could predict the quality accurately when only a small set of questions and their answers are labeled. To further validate the capability of \cop\ to deal with the sparsity problem, we consider the case when the labels of questions and answers are not simultaneously available. That is, we first fix 10\% of questions as well as their associated answers as the training set, and then delete the available answer scores. We compare the results with the {\em Separate} method in Fig.~\ref{F:Q10deleteA}. As we can see from Fig.~\ref{F:Q10deleteA-a}, our four methods are all better than the {\em Separate} method in terms of the prediction error. For example, in the extreme case when there are no answer scores available (i.e., 0\% answers), the {\em Separate} method can only guess the labels randomly, while our \copqq\ method can still predict the quality with prediction error less than 0.35. As to the comparison among our four methods, the \copqq\ method performs better than others when there are no answer scores available, and the four methods become close to each other wrt prediction error as the available answer scores increase. In Fig.~\ref{F:Q10deleteA-q}, we also show the corresponding prediction error for questions. As we can see, in all cases, our methods are better or at least close to the {\em Separate} method in terms of prediction error.

Similarly, we also choose 10\% of questions and their associated answers as the training set, and then delete the available question scores. The results are shown in Fig.~\ref{F:Q10deleteQ}. Again, as we can see from Fig.~\ref{F:Q10deleteQ-q}, our methods are better wrt prediction error when the available question labels are sparse. The corresponding answer quality prediction results are also shown in Fig.~\ref{F:Q10deleteQ-a}, and our four methods are substantially better than the {\em Separate} method wrt prediction error.

\subsection{Efficiency Results}

\begin{figure}[t]
  \centering
    \includegraphics[width=2.5in]{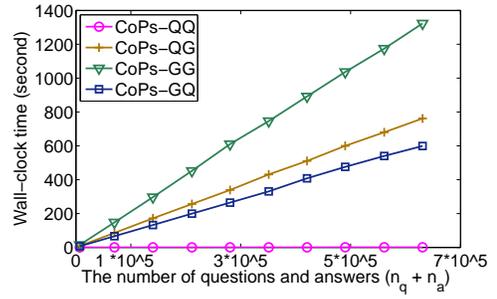}
  \caption{The wall-clock time of our methods. Our methods scale linearly wrt the data size ($n_q + n_a$).}
\label{F:efficiency}
\end{figure}

\begin{figure}[t]
  \centering
  \subfigure[Question quality prediction]{
  \label{F:ratio-q}\centering
    \includegraphics[width=1.5in]{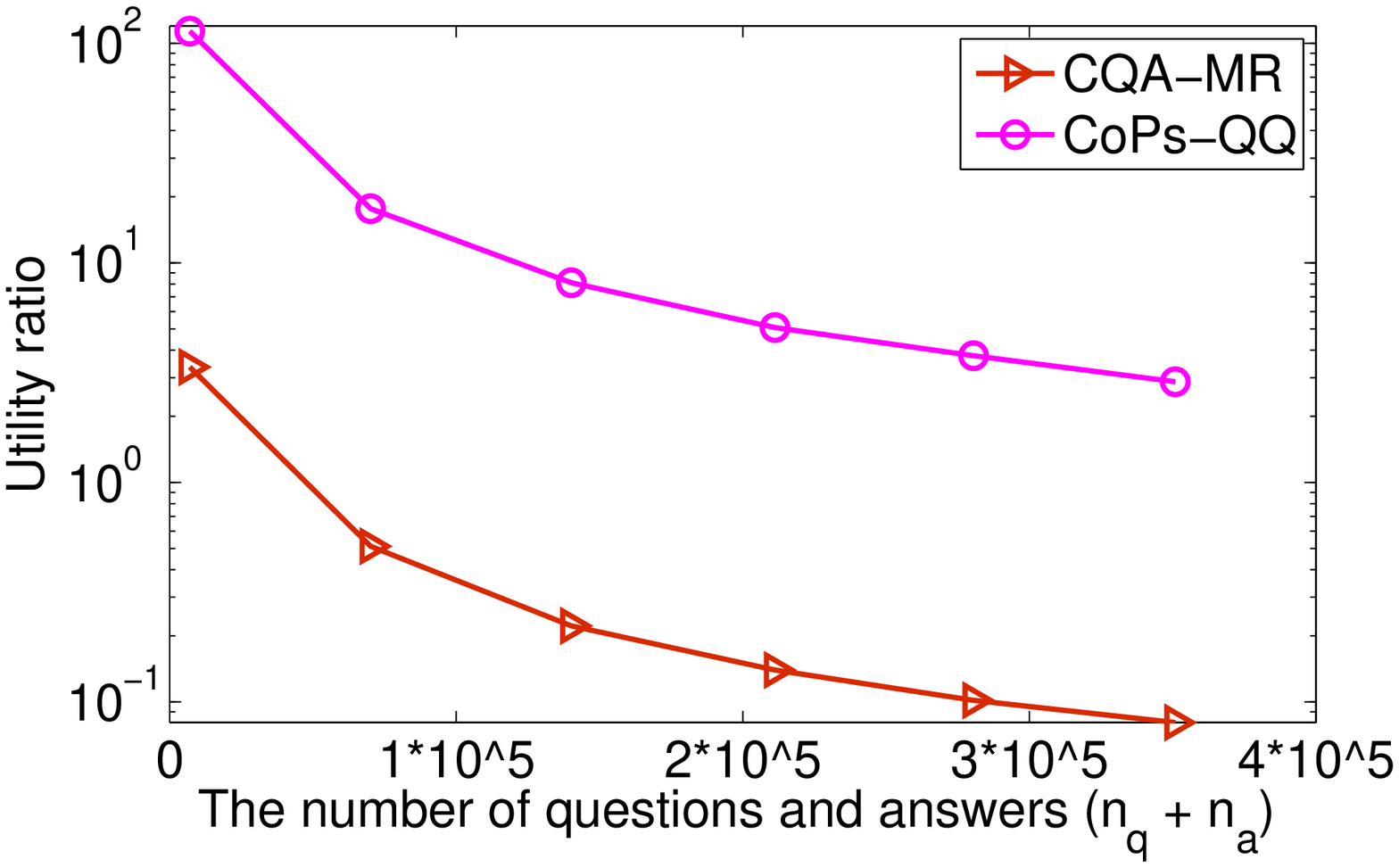}}
  \hspace{0.1in}
  \subfigure[Answer quality prediction]{
  \label{F:ratio-a}\centering
    \includegraphics[width=1.5in]{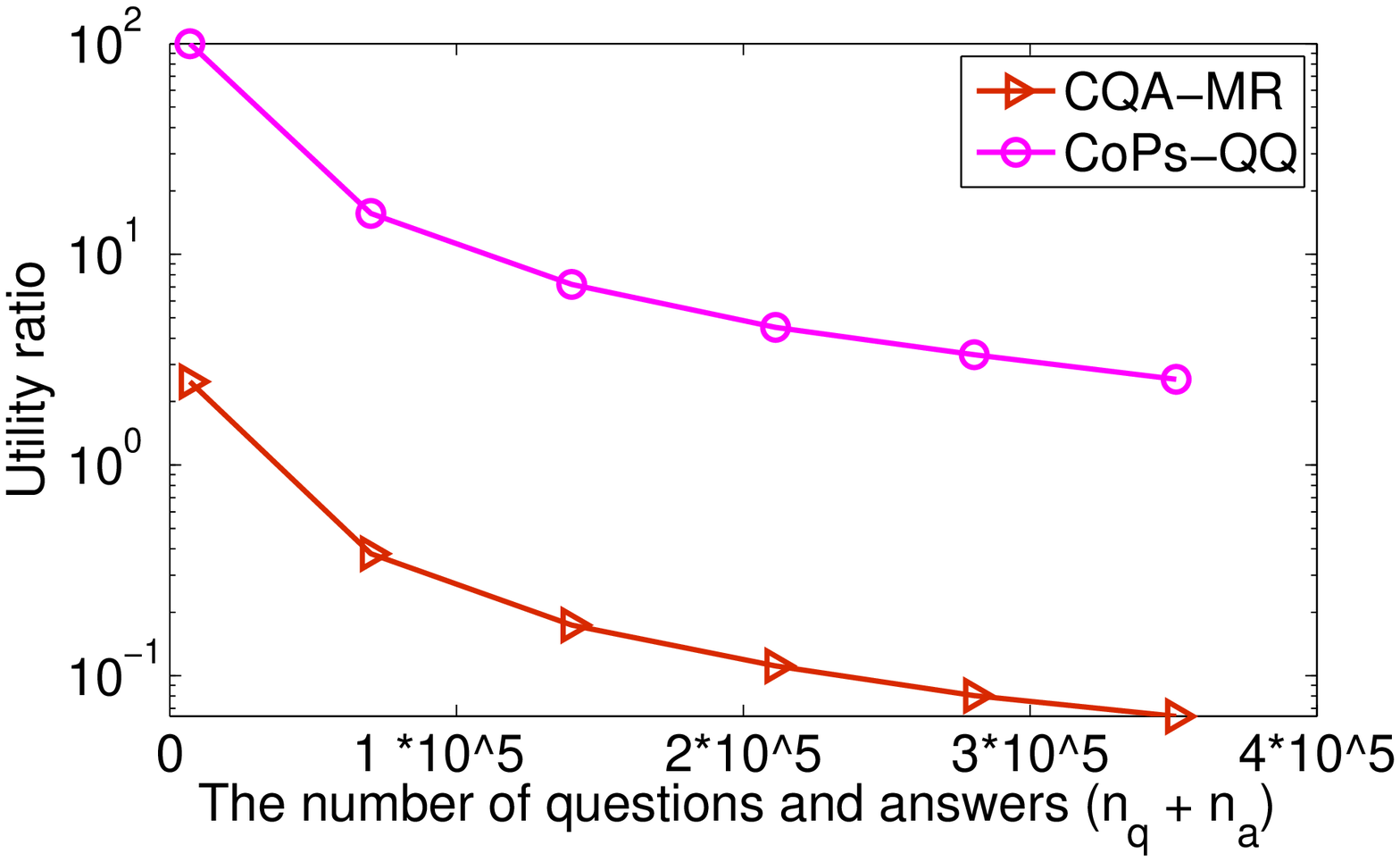}}
  \caption{The utility ratio of \copqq\ and the {\em CQA-MR} method. \copqq\ has a higher utility ratio.}
\label{F:ratio}
\end{figure}

Here, we also study the efficiency of \cop\ by recording the wall-clock time in the training step.
First, we vary the size of the training set, and plot the training time against the number of the questions and answers in the training set. The results of our four methods (i.e., \copqq\, \copqg\, \copgg\, \copgq) are shown in Fig.~\ref{F:efficiency}. As we can see, all our four methods scale linearly wrt the size of the training set, which is consistent to the algorithm analysis in Lemma~\ref{L:efficiency}. Notice that the \copqq\ method is much faster than the other three methods because we could derive the closed-form solution for it.

Next, we consider to combine effectiveness with efficiency, and define the utility ratio as $(1 - prediction~error) / wall-clock~time$. This utility ratio reflects the prediction accuracy in a given time period, and the higher the utility ratio the better the method. We compare the utility ratio of the \copqq\ method with that of the {\em CQA-MR} method for both question prediction and answer prediction in Fig.~\ref{F:ratio}. As we can see, \copqq\ performs better in both question prediction and answer prediction. Overall, among all the four variants of \cop, \copqq\ has the similar accuracy as others, while runs faster. Based on these efficiency results, together with the effectiveness results, we recommend \copqq\ in practice.

\subsection{Parameter Sensitivity}

\begin{figure}[t]
  \centering
  \subfigure[Question quality prediction vs. $\eta$]{
    \includegraphics[width=1.5in]{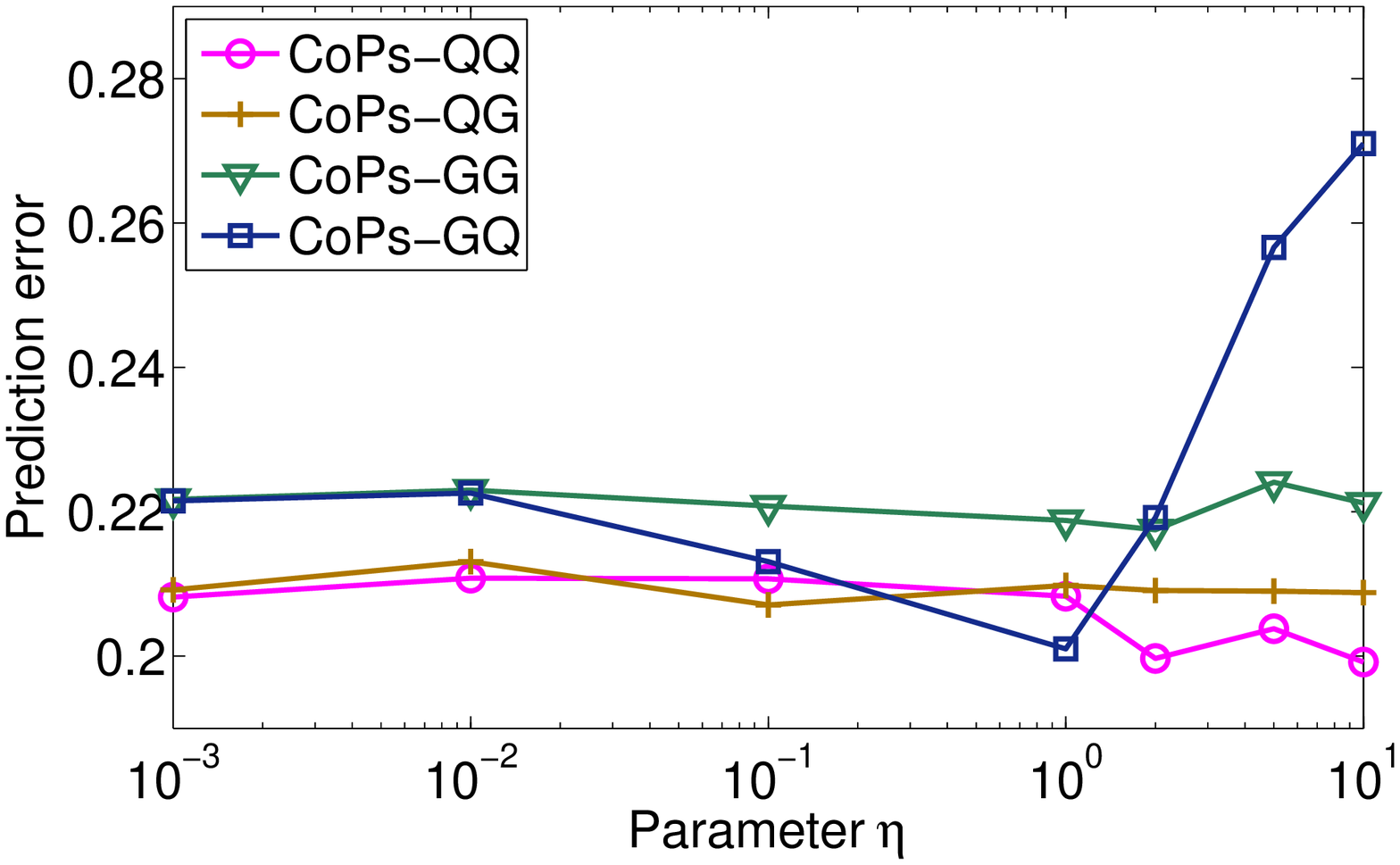}}
  \hspace{0.1in}
  \subfigure[Answer quality prediction vs. $\eta$]{
    \includegraphics[width=1.5in]{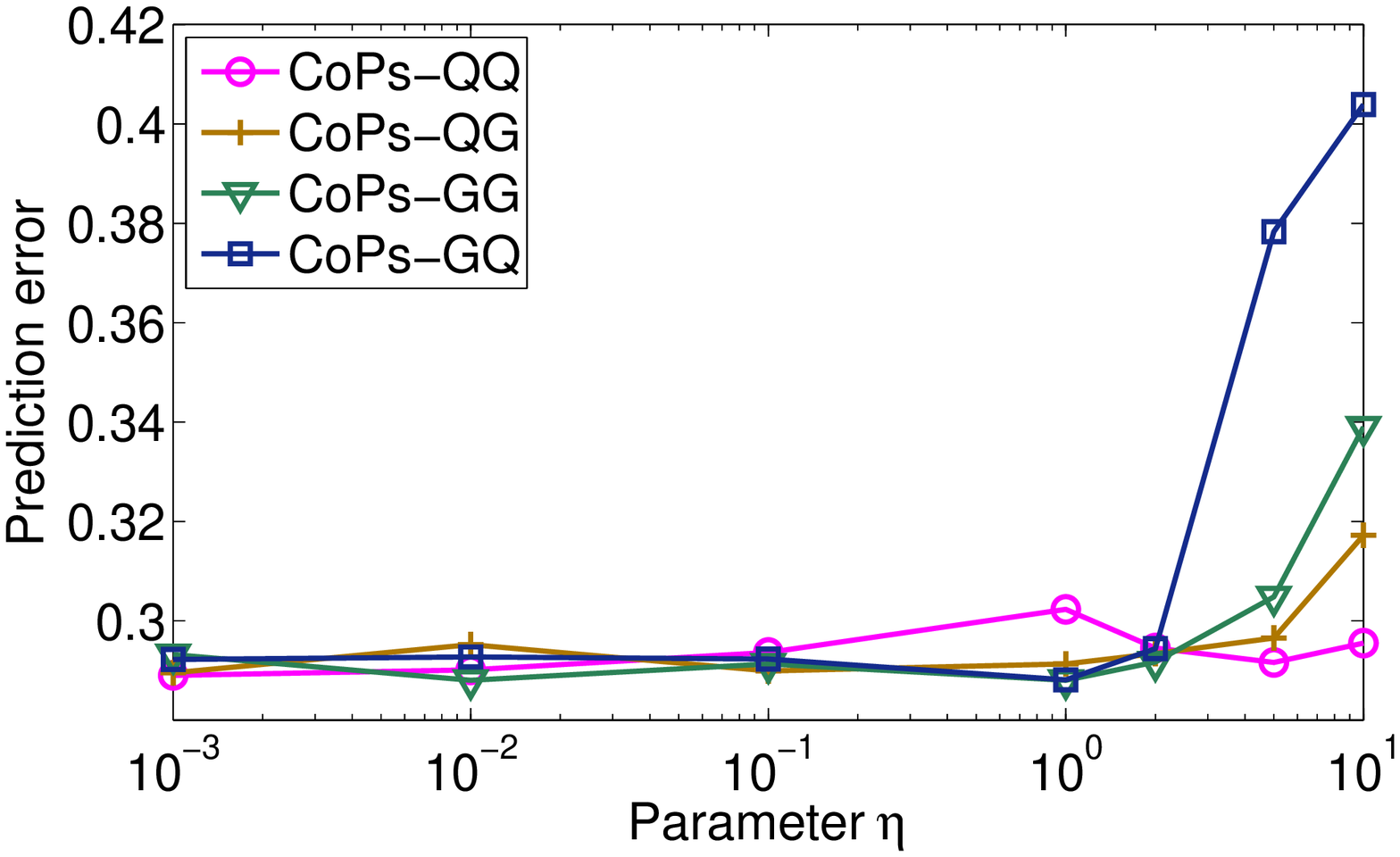}}
  \hspace{0.1in}
  \subfigure[Question quality prediction vs. $\lambda$]{
    \includegraphics[width=1.5in]{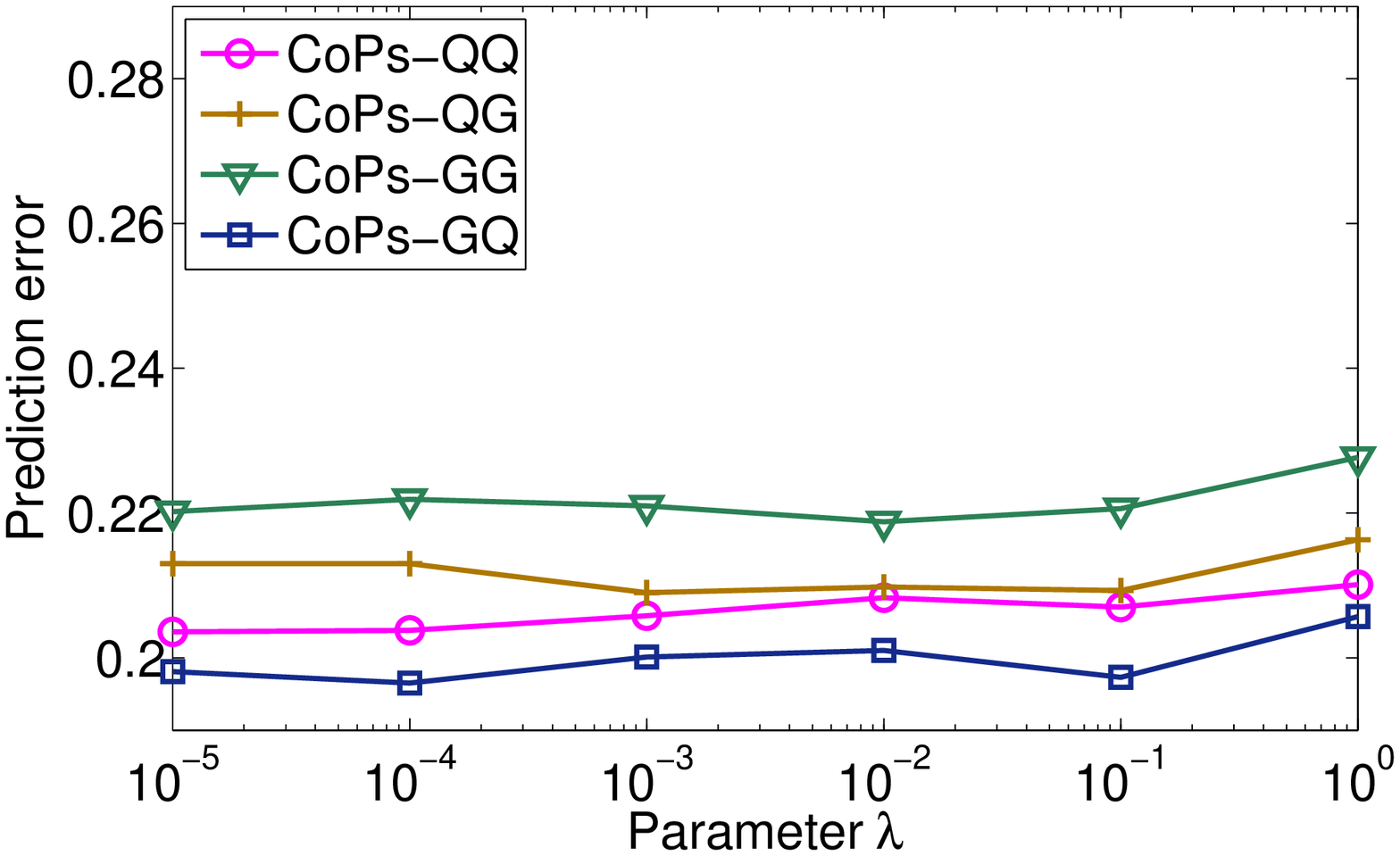}}
  \hspace{0.1in}
  \subfigure[Answer quality prediction vs. $\lambda$]{
    \includegraphics[width=1.5in]{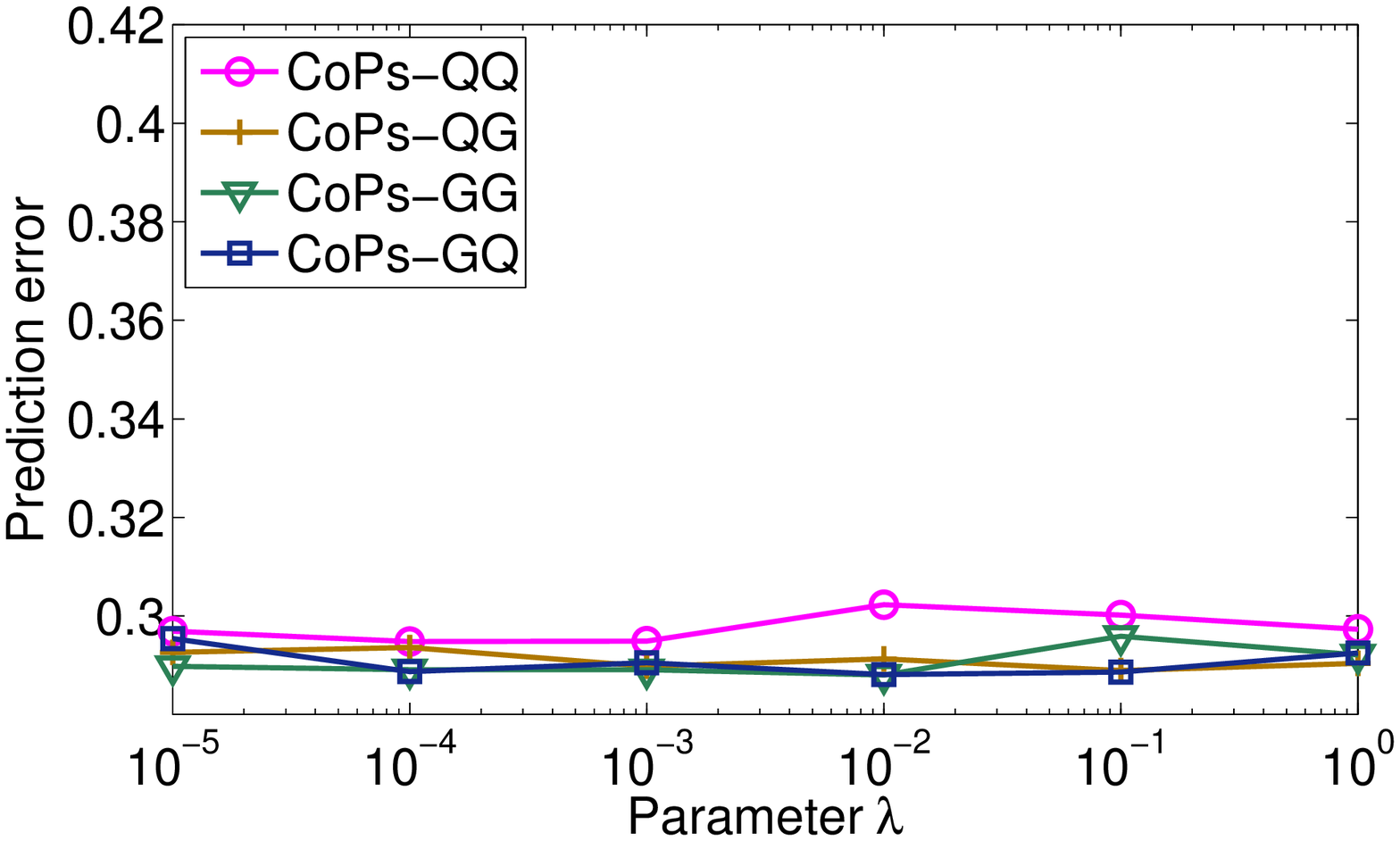}}
  \caption{Parameter sensitivity experiment. Our methods are robust wrt both parameters in a wide range.}
\label{F:pamameter}
\end{figure}

Finally, we test the parameter sensitivity of \cop. There are two parameters in Eq.\eqref{E:co-formulation}, i.e., $\eta$ which indicates the importance of the quality correlation term, and $\lambda$ which controls the amount of regularization. The results are shown in Fig.~\ref{F:pamameter}. As we can see, the prediction error of our methods stays stable when $\eta \leqslant 1$, which means that we may still have good prediction results when we put the same emphasis on the quality correlation term compared to the previous two terms in the equation. For simplicity, We fix $\eta = 1$ in our experiments. As for $\lambda$, the prediction accuracy is stable over a large range $[10^{-5}, 10^{0}]$, and we fix it to 0.01. Based on these results, we conclude that our methods are robust wrt the two parameters in a wide range.

\section{Related Work}
In this section, we briefly review related work including question/answer quality prediction, empirical studies on Stack Overflow, etc.

\textbf{Question/Answer Quality Prediction:} Question answering websites have become valuable knowledge bases which receive millions of visits and queries each day. As a result, several methods have been proposed to identify relevant questions for a given query (e.g.~\cite{zhou2011learning}). To further improve the usefulness of the returned questions, the quality of these questions should also be considered. For example, Song et al.~\cite{song2008question} define question quality as the likelihood that a question is repeatedly asked by people, and evaluate such measurement in the setting of question search. In addition to re-ranking the returned questions for a given query, question quality can be used to recommend questions to prominent places so that users can easily discover them~\cite{sun2009learning}.

Similar to question quality prediction, answer quality prediction could also be used to directly identify high-quality answers for a given user query. For example, Jeon et al.~\cite{jeon2006framework} and Suryanto et al.~\cite{suryanto2009quality} use human annotators to label the quality of the answers, and evaluate the usefulness of answer quality by incorporating it to improve retrieval performance.

The main focus on the above work is to employ the quality of questions/answers so as to improve the performance of information retrieval. In contrast, we observe the positive quality correlation between questions and their answers, and propose to jointly predict the quality of questions and answers by leveraging such correlation. As a result, on one hand, we can predict the quality of questions/answers more accurately; and on the other hand, we can exploit the usage of question-answer pairs.

Agichtein et al.~\cite{agichtein2008finding} and Bian et al.~\cite{bian2009learning} also aim to predict the quality of both questions and answers. Their focus is to tackle the sparsity problem where only a small number of questions/answers are labeled. As shown in our experiments, our method can also deal with the sparsity problem by leveraging the quality correlation between questions and their answers. In the methodology aspect, Agichtein et al.~\cite{agichtein2008finding} still treat question quality prediction and answer quality prediction as separated problems. For the method proposed by Bian et al.~\cite{bian2009learning}, as pointed in section 5.1, there are two important differences between their method ({\em CQA-MR}) and the our \cop, which lead to significant performance difference (See Fig.~\ref{F:QAfixK} for an example).

\textbf{Quality Measurement:} There are several types of measurement to quantify the quality of questions and answers. First, Liu et al.~\cite{liu2008predicting} propose to measure the questioner satisfaction, i.e., which answer the questioner will probably choose as the accepted answer. Later on, this problem is followed up by several researchers~\cite{shah2010evaluating,adamic2008knowledge}. However, accepted answers are not necessarily the highest-quality answers due to timing and subjectivity issues.

\hide{Liu et al.~\cite{liu2008predicting} first propose the questioner satisfaction prediction in CQA. They find that question features (e.g., length of the subject, posting time, etc.) and questioner history information are very useful for the prediction task.
Shah and Pomerantz~\cite{shah2010evaluating} also predict which answers the questioner would be satisfied with. They extract a set of features from the data and found that these features worked better than human labeling on several quality aspects in terms of questioner satisfaction prediction.
Adamic et al.~\cite{adamic2008knowledge} studied the forum categories and user behavior patterns in Yahoo! Answer. They also formulated a classification problem to predict the acceptance of answers based on the features from the question and the answerer.
All the above work considers the prediction of accepted answers. However, accepted answers are not necessarily the highest-quality/highest-score answers due to timing and subjectivity issues. Different from the above work, we aim to predict the quality of question and answers which is voted and determined by the whole community.}

To overcome the subjectivity issue of questioner satisfaction, many proposals resort to the quality measures derived from long-term community voting or human labeling.
For example, Harper et al.~\cite{harper2008predictors} conduct a field study on several question answering websites to seek the reasons for high-quality answers. They use the human labels as the quality indicator and find that factors such as community effect and payment play important roles in answer quality while rhetorical strategy and question type have little effect.
Nahehi et al.~\cite{nasehi2012makes} study the answer quality of code examples in Stack Overflow. They use the community voted score of an answer as the quality measure. Similar to this work, we also use the voted score, which is the difference between the number of up-votes and down-votes from the community, as the quality measure.

\textbf{Empirical Study on Stack Overflow:} Due to the great value of CQA in helping software development, many empirical studies are conducted on Stack Overflow. For example, Treude et al.~\cite{treude2011programmers} investigate the website to identify which types of questions are frequently asked and answered by the programmers. Parnin et al.~\cite{parnin2012crowd} study whether Stack Overflow can be used as a substitute of API documentation. Barua et al.~\cite{barua2012developers} analyze the text content of the posts in Stack Overflow to discover the current hot topics that software developers are discussing. Mamykina et al.~\cite{mamykina2011design} try to find the success design choices of Stack Overflow so that the lessons can be reused for other applications. One of the main reasons that Mamykina et al. found is the tight involvement of founders and moderators in the community. Our work could be used to automatically support their moderation by identifying the high-quality and low-quality posts in their early stage. In summary, different from the existing empirical studies, our focus is on the quality of questions and answers in Stack Overflow, and such post quality is essential for the reuse of CQA knowledge.

\textbf{Other Related Work:} There are some other recent focuses that are potentially related to our work. For example, Tausczik and Pennebaker~\cite{tausczik2011predicting} empirically study the correlation between user reputation and post quality in MathOverflow, and find that both offline and online reputation points are good predictors for post quality.
Gottipati et al.~\cite{gottipati2011finding} focus on how to find relevant answers in software forum when there could be many answers for a single question.
Kumar et al.~\cite{kumar2010evolution} reveal the mutual effect between question and answer dynamics in Stack Overflow, and prove that certain equilibrium can be achieved from a theoretical perspective.
Liu et al.~\cite{liu2011predicting} propose the problem of CQA searcher satisfaction, i.e., will the answer in a CQA satisfies the information searcher using the search engines. They divide the searcher satisfaction problem into three subproblems (i.e., query clarity, query-question match, and answer quality) and conclude that more intelligent prediction of answer quality is still in need.
How to route the right question to the right answerer~\cite{zhou2009routing,li2010g,horowitz2010anatomy}, and how to predict the long-lasting value (i.e., the page views of a question and its answers)~\cite{anderson2012discovering}, are also studied by several researchers.

\hide{Anderson et al.~\cite{anderson2012discovering} put their focus on the community processes in Stack Overflow, and showed how the community processes could be used to identify the threads with long-lasting value and the threads that are in need of additional help. Our quality prediction could be the input of their work, as threads with high-quality question and high-quality answers might be of long-lasting value, and threads with high-quality question and low-quality answers might need more help.}

\section{Discussions and Future Work}
There are several issues we need to clarify and discuss in this work.

First, although to some extent it is intuitive that the quality of questions and that of their answers are correlated, the empirical verification of this intuition is missing. This is probably due to the fact that most CQA websites do not contain suitable quality measures for both questions and answers. As a result, the state-of-the-art methods mainly use human annotators to label a small set of questions/answers, and then train a model based on these labels to predict the quality of other questions/answers. We would like to point out that, even in those cases where human annotation is used, the quality correlation between questions and their answers could still be used to improve the prediction accuracy.
In addition, we have shown in our experiments that our method can tackle the sparsity problem when only a small number of human-annotated labels are available. Therefore, our method is applicable and helpful in many quality prediction tasks in CQA websites.

Second, in the context of Stack Overflow, predicting the quality of question-answer pairs in their early stage is also important. For example, high-quality question and high-quality answer pairs can be highlighted to attract more attention, high-quality question and low-quality answer pairs can be recommended to experts, and low-quality question and low-quality answer pairs can be considered for community moderation. In this work, we fix the time window of the early stage to 24 hours. In the future work, we plan to change the time window to 3 hours, 1 hour, or even the time before any answers have been posted.

Another direction of future work is to investigate the evolution between question quality and answer quality. It would be interesting to study the interplay between question quality and answer quality over time, and it is also interesting to employ this dynamics to further improve the accuracy of quality prediction.

\section{Conclusions}
In this paper, we study the relationship between the quality of questions and answers in the CQA setting. We start with an empirical study of the Stack Overflow dataset where we observe a strong quality correlation between questions and their associated answers. Armed with this observation, we next propose a family of algorithms to jointly predict the quality of questions and answers, for both quantifying numerical quality scores and differentiating the high-quality questions/answers from those of low quality. Extensive experimental evaluations show that our methods outperform the state-of-the-art methods in effectiveness, and scale linearly wrt the number of questions and answers.

\bibliographystyle{abbrv}
\bibliography{SEformal}

\appendix
Here, we present the detailed algorithms for Alg.~\ref{A:co-gd}.

The core of the algorithm is to iteratively update the coefficients (step~6-7):
\begin{eqnarray}\label{E:co-gd}
\boldsymbol\beta_q & \leftarrow & \boldsymbol\beta_q - \gamma  \frac{\partial{\mathcal{L}}}{\partial{\boldsymbol\beta_{q}}} \nonumber\\
\boldsymbol\beta_a & \leftarrow & \boldsymbol\beta_a - \gamma  \frac{\partial{\mathcal{L}}}{\partial{\boldsymbol\beta_{a}}}
\end{eqnarray}
where $\gamma$ is the learning step size, and the partial derivatives can be generally computed as:
\begin{eqnarray}\label{E:co-gd2}
\frac{\partial{\mathcal{L}}}{\partial{\boldsymbol\beta_{q}}} &=& \sum_{i=1}^{n_q} \frac{\partial{g(\mat{X}_{\tilde{q}}(i,:) \boldsymbol\beta_q, \mat{y}_q(i))}}{\partial{\boldsymbol\beta_{q}}} + \eta \sum_{i=1}^{n_q} \frac{\partial{h(\mat{X}_{\tilde{q}}(i,:) \boldsymbol\beta_q, \tilde{\mat{M}}(i,:) \mat{X}_{\tilde{a}} \boldsymbol\beta_a)}}{\partial{\boldsymbol\beta_{q}}} \nonumber\\
& & + 2 \lambda \boldsymbol\beta_{q} \nonumber\\
\frac{\partial{\mathcal{L}}}{\partial{\boldsymbol\beta_{a}}} &=& \sum_{j=1}^{n_a} \frac{\partial{g(\mat{X}_{\tilde{a}}(j,:) \boldsymbol\beta_a, \mat{y}_a(j))}}{\partial{\boldsymbol\beta_{a}}} + \eta \sum_{i=1}^{n_q} \frac{\partial{h(\mat{X}_{\tilde{q}}(i,:) \boldsymbol\beta_q, \tilde{\mat{M}}(i,:) \mat{X}_{\tilde{a}} \boldsymbol\beta_a)}}{\partial{\boldsymbol\beta_{a}}} \nonumber\\
& & + 2 \lambda \boldsymbol\beta_{a}
\end{eqnarray}

Notice that we have four variants based on the combination of loss functions, and the partial derivatives for each variants can be computed as follows:
\begin{eqnarray}
\frac{\partial{g_{square}}}{\partial{\boldsymbol\beta_{q}}} &=& 2 (\mat{X}_{\tilde{q}}(i,:) \boldsymbol\beta_{q} - \mat{y}_{q}(i)) \mat{X}_{\tilde{q}}(i,:)' \nonumber\\
\frac{\partial{g_{sigmoid}}}{\partial{\boldsymbol\beta_{q}}} &=& - (\frac{1}{1 + \exp(\mat{X}_{\tilde{q}}(i,:) \boldsymbol\beta_{q} \mat{y}_{q}(i))})^2 \exp(\mat{X}_{\tilde{q}}(i,:) \boldsymbol\beta_{q} \mat{y}_{q}(i)) \nonumber\\
 & & \mat{y}_{q}(i) \mat{X}_{\tilde{q}}(i,:)' \nonumber\\
\frac{\partial{g_{square}}}{\partial{\boldsymbol\beta_{a}}} &=& 2 (\mat{X}_{\tilde{a}}(j,:) \boldsymbol\beta_{a} - \mat{y}_{a}(j)) \mat{X}_{\tilde{a}}(j,:)' \nonumber\\
\frac{\partial{g_{sigmoid}}}{\partial{\boldsymbol\beta_{a}}} &=& - (\frac{1}{1 + \exp(\mat{X}_{\tilde{a}}(j,:) \boldsymbol\beta_{a} \mat{y}_{a}(j))})^2 \exp(\mat{X}_{\tilde{a}}(j,:) \boldsymbol\beta_{a} \mat{y}_{a}(j)) \nonumber\\
 & & \mat{y}_{a}(j) \mat{X}_{\tilde{a}}(j,:)' \nonumber\\
\frac{\partial{h_{square}}}{\partial{\boldsymbol\beta_{q}}} &=& 2 (\mat{X}_{\tilde{q}}(i,:) \boldsymbol\beta_{q} - \tilde{\mat{M}}(i,:) \mat{X}_{\tilde{a}} \boldsymbol\beta_{a}) \mat{X}_{\tilde{q}}(i,:)' \nonumber\\
\frac{\partial{h_{sigmoid}}}{\partial{\boldsymbol\beta_{q}}} &=&  - (\frac{1}{1 + \exp(\mat{X}_{\tilde{q}}(i,:) \boldsymbol\beta_{q} \tilde{\mat{M}}(i,:) \mat{X}_{\tilde{a}} \boldsymbol\beta_{a})})^2 \exp(\mat{X}_{\tilde{q}}(i,:) \boldsymbol\beta_{q} \nonumber\\
 & & \tilde{\mat{M}}(i,:) \mat{X}_{\tilde{a}} \boldsymbol\beta_{a}) \tilde{\mat{M}}(i,:) \mat{X}_{\tilde{a}} \boldsymbol\beta_{a} \mat{X}_{\tilde{q}}(i,:)' \nonumber\\
\frac{\partial{h_{square}}}{\partial{\boldsymbol\beta_{a}}} &=& 2 (\mat{X}_{\tilde{q}}(i,:) \boldsymbol\beta_{q} - \tilde{\mat{M}}(i,:) \mat{X}_{\tilde{a}} \boldsymbol\beta_{a}) \mat{X}_{\tilde{a}}' \tilde{\mat{M}}(i,:)' \nonumber\\
\frac{\partial{h_{sigmoid}}}{\partial{\boldsymbol\beta_{a}}} &=&  - (\frac{1}{1 + \exp(\mat{X}_{\tilde{q}}(i,:) \boldsymbol\beta_{q} \tilde{\mat{M}}(i,:) \mat{X}_{\tilde{a}} \boldsymbol\beta_{a})})^2 \exp(\mat{X}_{\tilde{q}}(i,:) \boldsymbol\beta_{q} \nonumber\\
 & & \tilde{\mat{M}}(i,:) \mat{X}_{\tilde{a}} \boldsymbol\beta_{a}) \mat{X}_{\tilde{q}}(i,:) \boldsymbol\beta_{q} \mat{X}_{\tilde{a}}'\tilde{\mat{M}}(i,:)' \nonumber
\end{eqnarray}

\balance
\end{document}